\documentclass{raa}  
\usepackage{graphicx,times}
\usepackage{natbib}
\usepackage{amssymb,amsmath}
\bibpunct{(}{)}{;}{a}{}{,}

\usepackage[a4paper=true,dvipdfm=true,pagebackref=true]{hyperref}
\hypersetup{colorlinks = true, linkcolor = green, anchorcolor = red, citecolor = blue, filecolor = red, pagecolor = red, urlcolor = red}

\begin{document}

   \title{Modeling the High-energy Emission from the Gamma-ray Binary 1FGL J1018.6-5856
}

   \volnopage{Vol.0 (20xx) No.0, 000--000}      
   \setcounter{page}{1}                       

   \author{A. M. Chen \inst{1}
   \and C. W. Ng\inst{2}
   \and J. Takata \inst{1}
   \and Y. W. Yu \inst{3}
   }

   \institute{School of physics, Huazhong University of Science and Technology, Wuhan 430074, China {\it chensm@mails.ccnu.edu.cn; takata@hust.edu.cn; yuyw@mail.ccnu.edu.cn}\\
       \and Department of Physics, The University of Hong Kong, Pokfulam Road, Hong Kong, China\\
       \and College of Physical Science and Technology, Central China Normal University, Wuhan 430079, China\\
\vs\no
   {\small Received~~2020~~November 11; accepted~~2021~~March 14}}

\abstract{1FGL J1018.6-5856 is a high mass gamma-ray binary containing a compact object orbiting around a massive star with a period of 16.544 d. If the compact object is a pulsar, non-thermal emissions are likely produced by electrons accelerated at the termination shock, and may also originate from the magnetosphere and the un-shocked wind of the pulsar.
In this paper, we investigate the non-thermal emissions from the wind and the shock with different viewing geometries and study the multi-wavelength emissions from 1FGL J1018.6-5856. We present the analysis results of the \textit{Fermi}/LAT using nearly 10 years of data. The phase-resolved spectra indicate that the GeV emissions comprise a rather steady component that does not vary with orbital motion and a modulated component that shows flux maximum around inferior conjunction.
The keV/TeV light curves of 1FGL J1018.6-5856 also exhibit a sharp peak around inferior conjunction, which are attributed to the boosted emission from the shock, while the broad sinusoidal modulations could be originating from the deflected shock tail at a larger distance. The modulations of GeV flux are likely caused by the boosted synchrotron emission from the shock and the IC emission from the un-shocked pulsar wind, while the steady component comes from the outer-gap of the pulsar magnetosphere.
Finally, we discuss the similarities and differences of 1FGL J1018.6-5856 with other binaries, like LS 5039.
\keywords{binaries: close --- gamma rays: stars --- X-rays: binaries
--- radiation mechanisms: non-thermal}
}

   \authorrunning{A. M. Chen, C. W. Ng, J. Takata, \& Y. W. Yu }            
   \titlerunning{Modeling High-energy Emissions from 1FGL J1018.6-5856}  

   \maketitle

%
%
\section{Introduction}\label{sect:intro}   
Surveys with ground-based Cerenkov telescopes (e.g., \textit{H.E.S.S.}, \textit{MAGIC} and \textit{VERITAS}) and space-based satellites (e.g., \textit{Fermi}) have discovered a new class of binary systems that emit luminous $\gamma$-rays, which are called gamma-ray binaries.
These binaries are comprised of a stellar-mass compact object orbiting around a massive star which emit broadband emission with radiation power peaking in the $\gamma$-ray band (\citealt{Dubus2013}). The massive stars are type O or Be stars, while the compact objects can be neutron stars (NSs) or stellar-mass black holes (BHs).
There are two kinds of emission models being proposed for such a kind of binaries:
(1) in the micro-quasar scenario, the compact object accretes outflows or envelope matter from the companion star and launches a bipolar relativistic jet. Electrons in the jet up-scatter off the black-body photons from the companion or the synchrotron photons from the jet and then produce the observed emissions (e.g., \citealt{Bosch-Ramon2004a,Bosch-Ramon2004b,Bosch-Ramon2009});
(2) in the pulsar scenario: a termination shock would be formed by a collision between pulsar wind and stellar outflows, and shock accelerated electrons would radiate broadband emissions via inverse Compton (IC) scattering and synchrotron radiation (e.g., \citealt{Tavani1997}). Besides the shock radiation, the magnetospheric emission and IC scattering in the wind will also produce the observed $\gamma$-rays (\citealt{Kapala2010}).

The $\gamma$-ray source 1FGL J1018.6-5856 (hereafter J1018) was certified as a gamma-ray binary by \cite{Corbet2011} based on the blind search for periodic sources in the first \textit{Fermi}/LAT catalog.
The follow-up observations of the system in radio, optical and X-ray also confirmed the binary nature with a period of 16.6 d (\citealt{Fermi2012}).
The optical spectroscopy suggested that the massive companion is a type O6V((f)) star with a temperature of $T_{\star}\simeq38900\ \rm{K}$ and distance of $d_{\rm L}=5\pm2\rm{ kpc}$ from Earth (\citealt{Napoli2011}). Recently, a more accurate distance of J1018 was updated to $d_{\rm L}=6.4^{+1.7}_{-0.7}\rm{ kpc}$ by \cite{Marcote2018} based on observations from Australia Long Baseline Array, \textit{Gaia} Data Release 2 (DR2) and UCAC-4 catalog.

The X-ray observations of J1018 by \textit{NuSTAR}, \textit{XMM-Newton} and \textit{Swift/XRT} were presented in \cite{An2013,An2015}.
The X-ray light curves exhibit a periodic flare around phase 0\footnote{The $\gamma$-ray maximum is denoted as phase 0 in \cite{Fermi2012}, and we use the same denotation in this paper.} and a broad modulation component which peaks around 0.3-0.4. The X-ray spectrum can be fit well with a power-law function, which favors the shock interaction scenario rather than the accretion model (\citealt{An2015}).
The high-energy (HE) $\gamma$-rays detected by \textit{Fermi}/LAT show significant modulations in the luminosity and spectral shape (\citealt{Fermi2012,An2017}). The spectra around GeV band are characterized by a power-law with exponential cut-off (PLEC) function. \cite{An2017} found that the orbital variation in the lower energy $\gamma$-ray is similar to that of X-rays, while the $\gamma$-ray flux above 1 GeV changes significantly.
The \textit{H.E.S.S.} telescope also detected very-high-energy (VHE) $\gamma$-rays from J1018 (\citealt{HESS2012,HESS2015}). The TeV light curve also displays a similar behavior as the X-rays, which also exhibits a flux maximum at phase 0. The measured spectrum extends to above 20 TeV and the spectral shape indicates a modest influence from the $\gamma$-ray absorption (\citealt{HESS2015}).

Unfortunately, the nature of the compact object of J1018 is as yet unknown. Although the measured spectral shape by \textit{Fermi}/LAT is similar to gamma-ray pulsars, there is still no direct detection of a pulsed signal, and therefore an accreting NS or BH still cannot be discarded explicitly (\citealt{Fermi2012}).
\cite{Waisberg2015} presented a radial velocity (RV) measurement of J1018 with the Cerro Tololo Inter-American Observatory (CTIO) telescope. Their analysis showed a semi-amplitude modulation of $12-40 \ \rm{km\cdot s^{-1}}$, and indicated most likely a compact object mass with $M_{\rm X}<2.2M_{\odot}$.
\cite{Strader2015} performed further spectroscopy of the optical companion with the Southern Astrophysical Research (SOAR) telescope. The RV semi-amplitude was constrained to be $11-12\ \rm{km\cdot s^{-1}}$, which suggested an NS primary of the binary, although a BH can not be discarded if the inclination angle of the orbit is very small. They also found that both the X-ray and $\gamma$-ray maxima emissions occur at inferior conjunction (INFC).
A follow-up RV study of J1018 with the Southern African Large Telescope (SALT) observations was performed by \cite{Monageng2017}. Combining with previous RV studies, they obtained constraints on the eccentricity and the inclination angle of J1018 with $e=0.31\pm0.16$ and $i\geq26^{\circ}$, respectively, for an NS primary. Their study also suggested that the periastron phase of the compact object occurs around INFC (see Fig. 4 in \citealt{Monageng2017}).

With the growing evidence suggesting an NS primary, here we investigate HE emissions of J1018 under the pulsar scenario and attempt to constrain the properties of the NS. The paper is organized as follows. In Sect. \ref{sect:data}, we report our analysis results of J1018 with \textit{Fermi}/LAT. Then, we describe the emission model in Sect. \ref{sect:model} and compare our results with observational data in Sect. \ref{sect:results}. Finally, we summarize our work and discuss the similarities and differences of J1018 with other binaries in Sect. \ref{sect:summary}.

\section{Data Analysis}\label{sect:data}
\begin{figure}
\centering
\resizebox{0.70\textwidth}{!}{\includegraphics{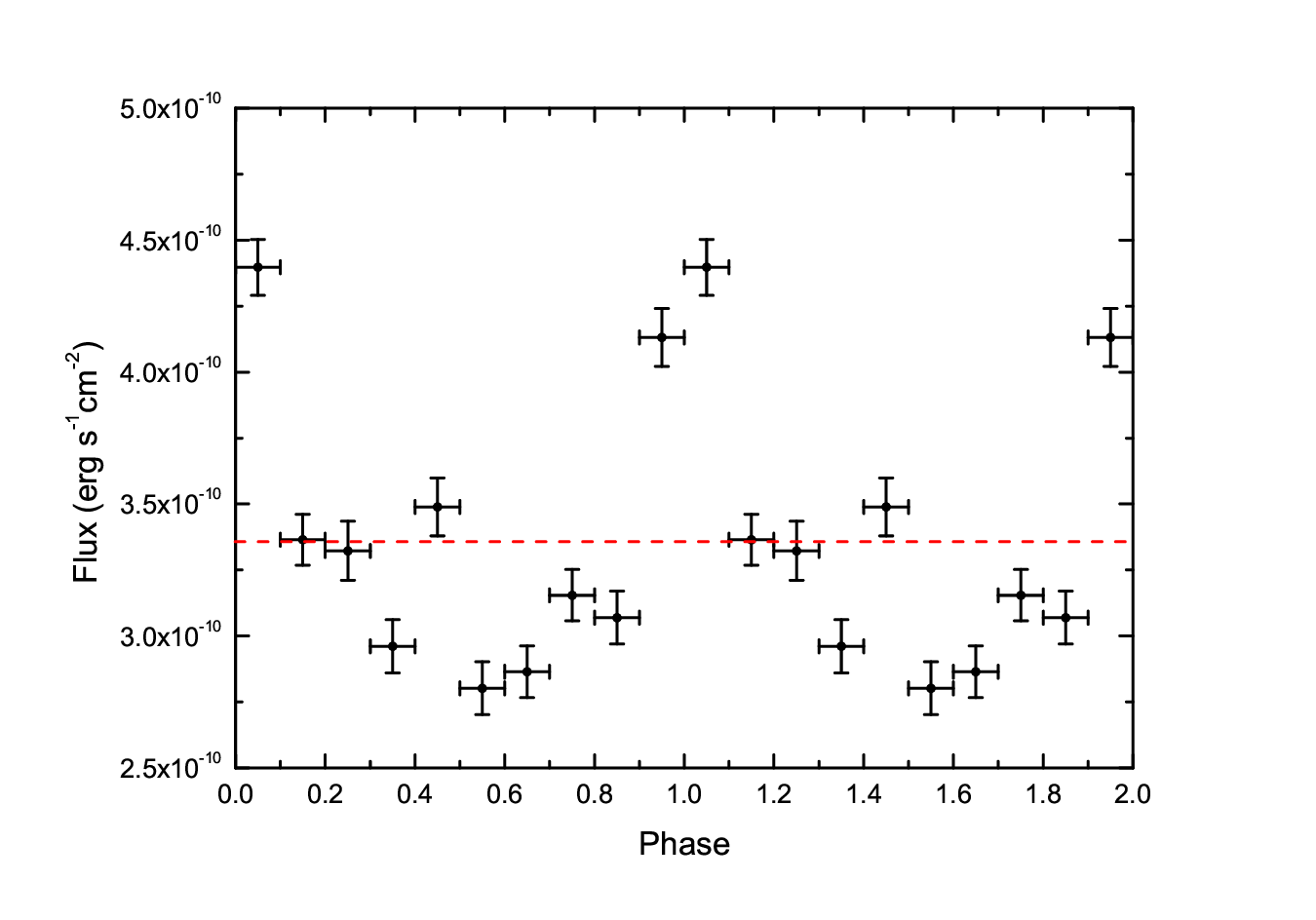}}
\caption{The orbital light curve of J1018 in 0.1 - 100
GeV obtained from binned likelihood analysis. The red dashed line
indicates the mean energy flux.}
\label{fig:LAT_LC}
\end{figure}

\begin{figure}
\centering
\resizebox{0.70\textwidth}{!}{\includegraphics{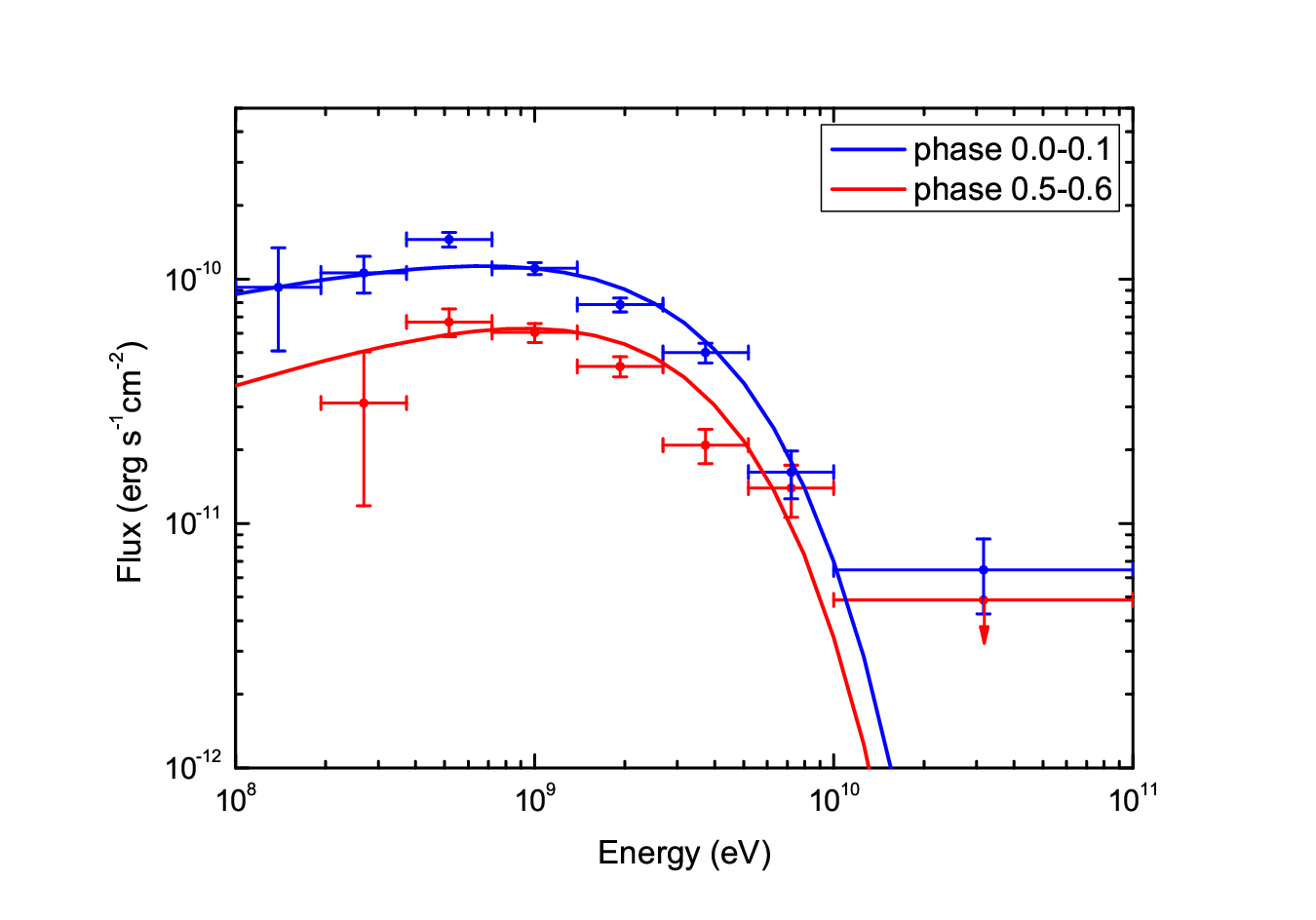}}
\caption{The orbital phase resolved spectra of J1018.
Red and blue curves are fitted by a PLEC function while the points with error bars are observational data and upper limit.}
\label{fig:LAT_SED}
\end{figure}

In this section, we analyze the HE emission of J1018 detected by \textit{Fermi}/LAT.
Photon events from 2008-August-09 to 2018-May-10 with energies of 0.1-100 GeV were selected from the ``Pass 8 Source" event class.
The region of interest (ROI) is a $20^\circ \times 20^\circ$ square centered at the epoch J2000 position of the source: $(\textrm{R.A.},\textrm{Dec})=(10^{\textrm{h}} 18^{\textrm{m}} 55.18^{\textrm{s}},-58^\circ 56^\prime 44.2^{\prime\prime})$.
We removed the events with zenith angle larger than $90^{\circ}$ to reduce contamination from the Earth's albedo.
The $gtlike$ tool was applied to perform maximum binned likelihood analysis to obtain the spectral models for all the 3FGL catalog sources that are within $25^{\circ}$ from the center of the ROI (gll\_psc\_v16.fit), the galactic diffuse emission (gll\_iem\_v06) and the isotropic diffuse emission (iso\_P8R2\_SOURCE\_V6\_v06) (\citealt{Acero2015}).
Four extended sources within the region: HESS J1303-631, Puppis A, Vela Jr and Vela X are modeled by the extended source templates of Fermi Science Support Center\footnote{\url{http://fermi.gsfc.nasa.gov/ssc/}}.
With the spectral indices fixed to the global fit and leaving only the normalization parameter free, we use the model to calculate the orbital flux. To get the orbital light curve of J1018, we fix the orbital period to be 16.544 d (\citealt{An2015}), then the TEMPO2 package (\citealt{Hobbs2006}) with Fermi plug-in (\citealt{Ray2011}) was utilized to assign an orbital phase for each event. The orbital light curve of J1018 is depicted in Fig. \ref{fig:LAT_LC}.
As we can see, the GeV emission displays significant orbital modulations, with the flux maxima around phase 0.0 and minimum around 0.5.
We perform spectral analysis in the selected phase intervals to investigate if the
spectrum of J1018 is varying throughout the orbital period. We defined the phase interval between 0.0-0.1 as high state and 0.5-0.6 as low state. We use the same data set described above and sub-selected these two states.
The spectral form of J1018 is modeled by a PLEC function
\begin{equation}\label{PLEC}
  \frac{dN}{dE}=N_0\left(\frac{E}{E_0}\right)^{-\Gamma_{\gamma}}\exp\left[-\left(\frac{E}{E_{\rm c}}\right)\right],
\end{equation}
with $N_0=5.65\times10^{-5}\ {\rm ph\cdot s^{-1}\cdot cm^{-2}\cdot erg^{-1}}, \Gamma_{\gamma}=1.82$, and $E_{\rm c}=2.96\ {\rm GeV}$ for high state, and $N_0=3.70\times10^{-5}
\ {\rm ph\cdot s^{-1}\cdot cm^{-2}\cdot erg^{-1}}, \Gamma_{\gamma}=1.64$, and $E_{\rm c}=2.42\ {\rm GeV}$ for low state.
The upper limits are derived when the detection significance is less than 3$\sigma$. The phase-resolved spectra are shown in Fig. \ref{fig:LAT_SED}. The spectra at these two states exhibit significant discrepancies at lower energy while the deviation become smaller at higher energy, which are consistent with the results of \cite{An2017}. The orbital modulation of GeV flux is similar to that of X-rays, manifesting maximum flux around INFC.

\section{Emission Model}\label{sect:model}
\begin{figure}
\centering
\resizebox{0.8\textwidth}{!}{\includegraphics{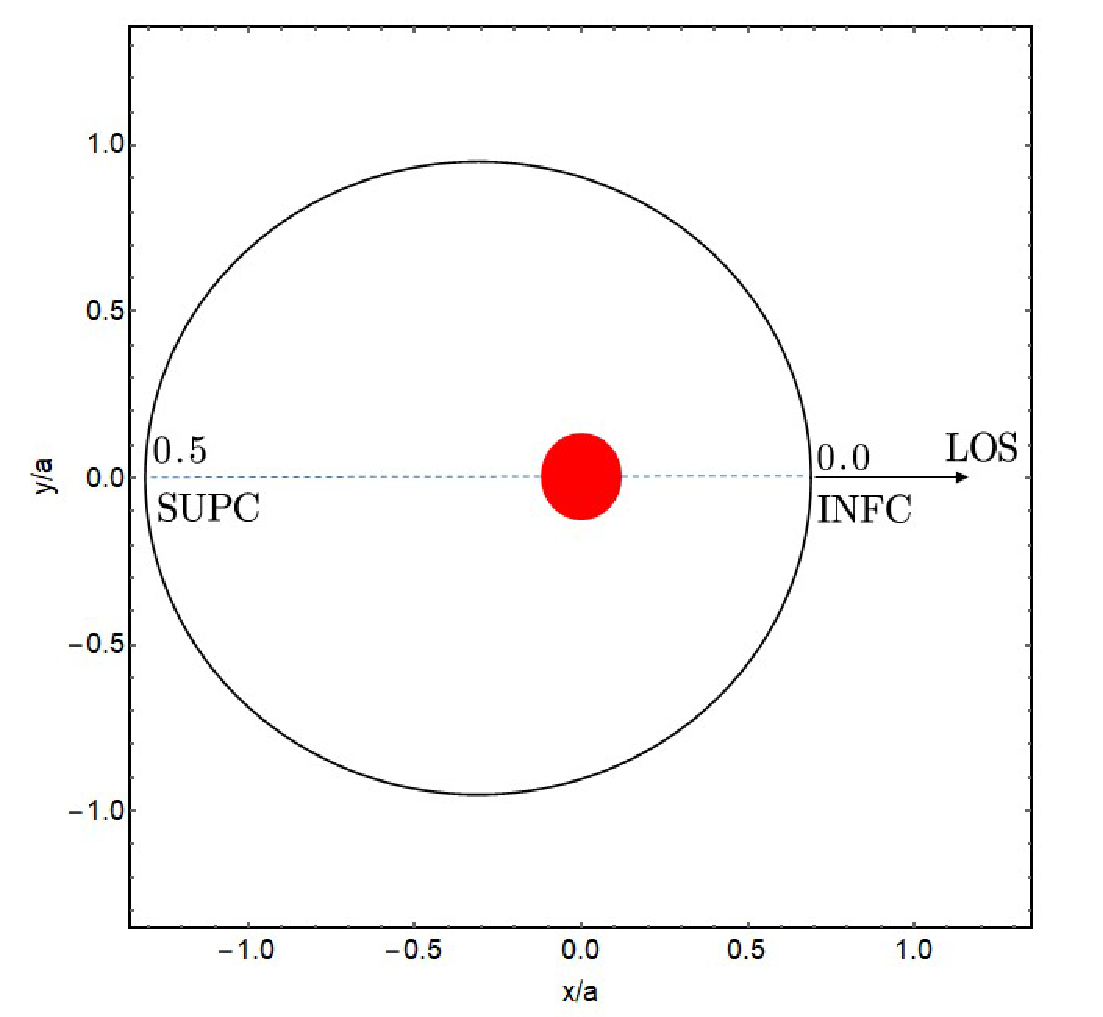}}
\caption{The orbit of J1018 with the parameters measured by \cite{Monageng2017}.
} \label{fig:orbit}
\end{figure}

\begin{figure}
\centering
\resizebox{0.8\textwidth}{!}{\includegraphics{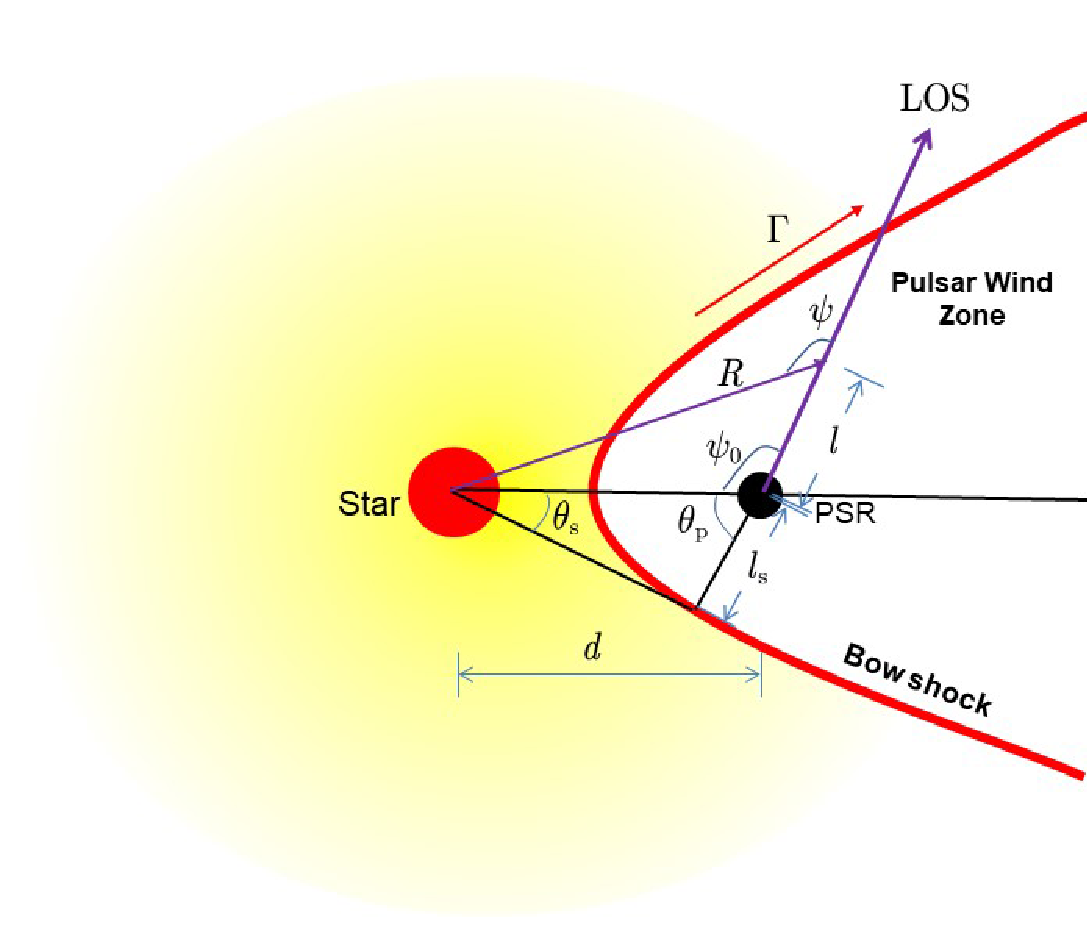}}
\caption{Geometry of the cold pulsar wind and the termination shock.
} \label{fig:shock}
\end{figure}

In this section, we describe the emission model for gamma-ray binaries under the pulsar scenario.
The pulsar wind is terminated by stellar outflows which produce an intra-binary bow shock (IBS). As the shocked flow propagates away from the apex due to the adiabatic expansion, the bulk Lorentz factor (LF) will increase gradually to mild relativistic velocity in the tail. The synchrotron radiation and IC scattering in the shock produce the observed X-rays and VHE $\gamma$-rays, respectively (\citealt{Dubus2006a,Chen2019}). Alternatively, the electrons in the pulsar wind zone (PWZ) will up-scatter the stellar photons to $\gamma$-rays, and the magnetospheric emission from the pulsar will also contribute to the observed emissions. The orbit and geometry of J1018 discussed in this paper are presented in Fig. \ref{fig:orbit} and \ref{fig:shock}, respectively. The related orbital parameters of J1018 are summarized in Table. \ref{table}.

\begin{table*}[t]
\caption{Parameters of J1018. \label{table}}
\begin{tabular}{l c c c}
\hline
\textbf{Parameter} &  \textbf{Symbol} & \textbf{Value}  & \textbf{Reference}\\
\hline
{\it System parameters} &  &  &\\
eccentricity  &  $e$  & $0.31\pm0.16$  & \citealt{Monageng2017}\\
orbital period   &$P_{\textrm{orb}}$  & 16.544 \textrm{days}& \citealt{An2015}\\
distance    &  $d_{\rm L}$   & $6.4^{+1.7}_{-0.7}$ \textrm{kpc} & \citealt{Marcote2018}\\
inclination angle of LOS  &$i$   & $45^{\circ}$ & Assumed$^\dag$ \\
true anomaly of LOS  & $\omega$ & $1\pm30^{\circ}$  & \citealt{Monageng2017}\\

\hline
{\it Pulsar and pulsar wind} &   &  & \\
spin-down power  & $L_{\textrm{sd}}$   &  $1\times 10^{36}\ {\rm erg\cdot s^{-1}}$   &Assumed$^\dag$  \\
rotation period  &  $P$  & 0.05 {\rm s}  & Assumed$^\dag$ \\
LF of pulsar wind at $r_{\rm L}$ & $\gamma_\textrm{L}$  & $1\times10^3$  & Assumed$^\dag$ \\
magnetization of pulsar wind at $r_{\rm L}$  & $\sigma_\textrm{L}$  & $1\times10^{2}$  & Assumed$^\dag$ \\

\hline
{\it Star and stellar outflows}  &  &  & \\
mass   & $M_\star$  & $22.9 M_\odot$ & \citealt{Monageng2017}\\
radius   & $R_\star$  & $9.3 R_\odot$ & \citealt{Monageng2017}\\
temperature  & $T_\star$  & $3.89\times10^4\ \textrm{K}$  & \citealt{Napoli2011}\\

\hline
{\it Termination shock} &  &  & \\
particle distribution index    & $p$   & $2.1$   & Assumed$^\dag$ \\
maximum LF of shocked flow   & $\Gamma$   & $2$  & Assumed$^\dag$ \\
\hline
\end{tabular}
\\
{{
\textbf{Notes.}\footnotesize $^\dag$ Model parameters. The values adopted in this table are chosen by modeling the observational data.
}}
\end{table*}

\subsection{Geometry of the termination shock}\label{sect:geometry}
The structure of the termination shock is decided by the dynamic balance between the relativistic pulsar wind and stellar outflows. Define the momentum flux ratio as
\begin{equation}\label{eta}
  \eta=\frac{L_{\rm{sd}}/c}{\dot{M}v_{\rm{w}}},
\end{equation}
where $L_{\rm sd}$ is the spin-down luminosity, $c$ is the speed of light, and $\dot{M}$ and $v_{\rm{w}}$ are the mass loss rate and the wind velocity of the massive star, respectively. Then the distance from the pulsar to the contact discontinuity of IBS is (\citealt{Canto1996, An2018})
\begin{equation}\label{rs}
  l_{\rm{s}}= d\frac{\sin\theta_{\rm s}}{\sin(\theta_{\rm p}+\theta_{\rm s})},
\end{equation}
with
\begin{equation}\label{theta}
  \theta_{\rm s}\cot\theta_{\rm s}= 1+\eta(\theta_{\rm p}\cot\theta_{\rm p}-1),
\end{equation}
where $d$ is the binary separation, and $\theta_{\rm s}$ and $\theta_{\rm p}$ are the angles of the point at the shock related to the line joining the star and the pulsar, respectively. The half opening-angle of IBS can be approximated with (\citealt{Eichler1993}):
\begin{equation}\label{thetas}
\theta_{\rm{sh}}=2.1(1-\overline{\eta}^{2/5}/4)\overline{\eta}^{1/3},
\end{equation}
with $\overline{\eta}=\min(\eta,\eta^{-1})$. For most high mass gamma-ray binaries, the stellar outflows are more powerful than that of the pulsar (i.e., $\eta<1$), which means that the shock would wrap around the pulsar.

\subsection{IC scattering in the cold pulsar wind}\label{sect:PW}
The rotational energy of pulsars is mainly released via relativistic winds composed of B-field and e-pairs (\citealt{Michel1969}). Initially, the pulsar wind is dominated by Poynting flux, and it is converted into kinetic energy as the wind is spreading away at a larger distance (\citealt{Aharonian2012}). The detailed mechanisms of the dissipation of Poynting flux and the acceleration of particles in pulsar wind are still unclear.
To describe the dynamics of pulsar wind, we introduce the so-called
magnetization parameter which is defined as the ratio of magnetic energy density to pair kinetic energy density in the wind.
Assuming that the magnetization of pulsar wind evolves with radial distance in the form of a power-law (\citealt{Contopoulos2002,Kong2011,Kong2012,Takata2017}):
\begin{equation}\label{sigma}
  \sigma(l)=\sigma_{\rm L}\left(\frac{l}{r_{\rm L}}\right)^{-\alpha_{\sigma}},
\end{equation}
according to the energy conservation law, the LF evolution of electrons in PWZ can be written as (e.g., \citealt{Chen2015})
\begin{equation}\label{gammaw}
  \gamma_{\rm w}(l)\simeq\gamma_{\rm L}\frac{1+\sigma_{\rm L}}{1+\sigma},
\end{equation}
where $\sigma_{\rm L}$ and $\gamma_{\rm L}$ are the magnetization parameter and LF of pulsar wind at light cylinder $r_{\rm L}$, and $\alpha_{\sigma}$ is order of unity.

The companion star provides a large number of soft photons that would be up-scattered to higher energies by the pulsar wind electrons. For a monochromatic energy electron with an LF of $\gamma_{\rm w}\sim10^4$ and a stellar photon with $\epsilon_0\sim2.82kT_{\star}$, the characteristic energy of the up-scattered photon is
\begin{equation}
  E_{\gamma}\simeq4\gamma_{\rm w}^2\epsilon_0
  \sim9.66\times10^8{\rm{eV}}\left(\frac{\gamma_{\rm w}}{10^4}\right)^2\left(\frac{T_{\star}}{10^4}\right),
\end{equation}
in the Thompson regime (i.e. $\gamma_{\rm w}kT_{\star}/m_{\rm e}c^2\ll1$), or
\begin{equation}
  E_{\gamma}\simeq\gamma_{\rm w}m_{\rm e}c^2\sim
  5.07\times10^9{\rm{eV}}\left(\frac{\gamma_{\rm w}}{10^4}\right),
\end{equation}
in the Klein-Nishina regime (i.e. $\gamma_{\rm w}kT_{\star}/m_{\rm e}c^2\gg1$), which is located in the energy band of \textit{Fermi}/LAT. It means that the modulations of $\gamma$-rays observed by \textit{Fermi}/LAT could be contributed by the IC emission in PWZ. Since the pulsar is orbiting around the massive star, the IC emission from the wind is highly anisotropic.
Assuming that electrons are moving radially in the wind, the observed $\gamma$-rays are produced by electrons that are moving in the same direction (\citealt{Khangulyan2011}).
The IC scattering power at the frequency of $\nu$ for a single electron is given by (\citealt{Aharonian1981}):
\begin{eqnarray}\label{PEIC}
P\left(\nu,\gamma_{\rm w}\right) &=& 3\sigma_{\rm T}\int \frac{\nu f(\nu_{\rm s})}{4\gamma_{\rm w}^2\nu_{\rm s}^2}H(\xi,b_{\theta}){\rm d}\nu_{\rm s}, \\
\nonumber  H\left(\xi,b_{\theta}\right) &=& 1+\frac{\xi^{2}}{2(1-\xi)}-\frac{2\xi}{b_{\theta}(1-\xi)} +\frac{2\xi^{2}}{b_{\theta}^{2}(1-\xi)^{2}},
\end{eqnarray}
where $\xi=h\nu/\gamma_{\mathrm { w } } m _ { \mathrm { e } } c ^ { 2 }$, and $b _ { \theta } = 2 ( 1 -  \cos \theta _ { \mathrm { SC } } ) \gamma _ { \mathrm { w } } h \nu_ { s } /  m _ { \mathrm { e } } c ^ { 2 }.$ Under the point source approximation, the flux density of stellar photons is expressed as $f(\nu_{\rm s})=\pi(R_{\star}/R)^2(2h\nu_{\rm s}^3/c^2)[1/(\exp(h\nu_{\rm s}/kT_{\star})-1)]$, and
the scattering angle is determined by $\theta_{\rm SC}=\pi-\psi$, with $\psi$ being related to the distance $l$ from the pulsar, expressed as
\begin{equation}
  \psi =\left\{ \begin{matrix}
	\tan ^{-1}\left( \frac{d\sin \psi _0}{d\cos \psi _0-l} \right)&{\rm for}\ l<d\cos \psi _0\\
	\pi+\tan ^{-1}\left( \frac{d\sin \psi _0}{d\cos \psi _0-l} \right)&{\rm for}\ l>d\cos \psi _0.\\
\end{matrix} \right.
\end{equation}
The distance from the star is defined by $R^2=d^{2}+l^2-2dl\cos \psi _0$ and $\cos\psi_0=-\sin i\cos(\phi-\omega)$, with $i$ being the inclination angle of the orbit, and $\phi$ and $\omega$ being the true anomaly angle of the pulsar and the line of sight (LOS) projected to the orbital plane, respectively. The number of electrons in PWZ per unit of length is given by (\citealt{Yi2017})
\begin{equation}
   N_{\rm e}(l,\gamma_{\rm w})=\frac{L_{\rm sd}}{4\pi\gamma_{\rm w}(1+\sigma)m_{\rm e}c^3}.
\end{equation}
Finally, the observed flux from PWZ can be obtained by integrating over LOS
\begin{equation}\label{F0}
  F(\nu)=\frac{1}{d_L^2}\int_0^{l_{\rm s}} N_{\rm e}(l,\gamma_{\rm w}) P(\nu,\gamma_{\rm w})
   {\rm{d}}l.
\end{equation}
Different from the case of a free expanding pulsar wind as investigated by \cite{Ball2000}, the presence of the termination shock will reduce the size of the cold pulsar wind, and thus affect the IC emissions (\citealt{Ball2001,Cerutti2008}), so in the calculation of Eq. (\ref{F0}), we integrate over the length of un-shocked wind region towards the observer from the pulsar to the shock contact discontinuity surface $l_{\rm s}$.

In summary, the emissions from the un-shocked wind are mainly determined by the following parameters (\citealt{Khangulyan2011,Khangulyan2012}):
(1) the LF of un-shocked electrons $\gamma_{\rm w}$,
(2) the flux density spectrum of the seed photons $f(\nu_{\rm s})$,
(3) the length of PWZ towards the observer $l_{\rm s}$, and
(4) the scattering angle between the incoming and up-scattered photon $\theta_{\rm SC}$.
The orbital motion of the pulsar around its companion leads to the modulations of the above parameters, and thus affects the wind emission.
It is necessary to point out that the IC process in PWZ would reduce the particle energies, which provide a drag of pulsar wind and reduce the LF of un-shocked electrons (\citealt{Ball2000}).
For simplicity, the effect of Compton-drag on the dynamics of pulsar wind is not considered here.
The rest of the parameters (i.e., $f(\nu_{\rm s})$, $l_{\rm s}$, and $\theta_{\rm SC}$) are mainly determined by the distance of the emitting region from the star, the shock structure and the viewing angles, which are governed by the binary separation, the momentum flux ratio and the viewing angles. Next, we will explore the effects of the above parameters on the IC emission from the un-shocked wind.

\begin{figure}
\begin{minipage}{0.45\linewidth}
  \centerline{\includegraphics[width=8.0cm]{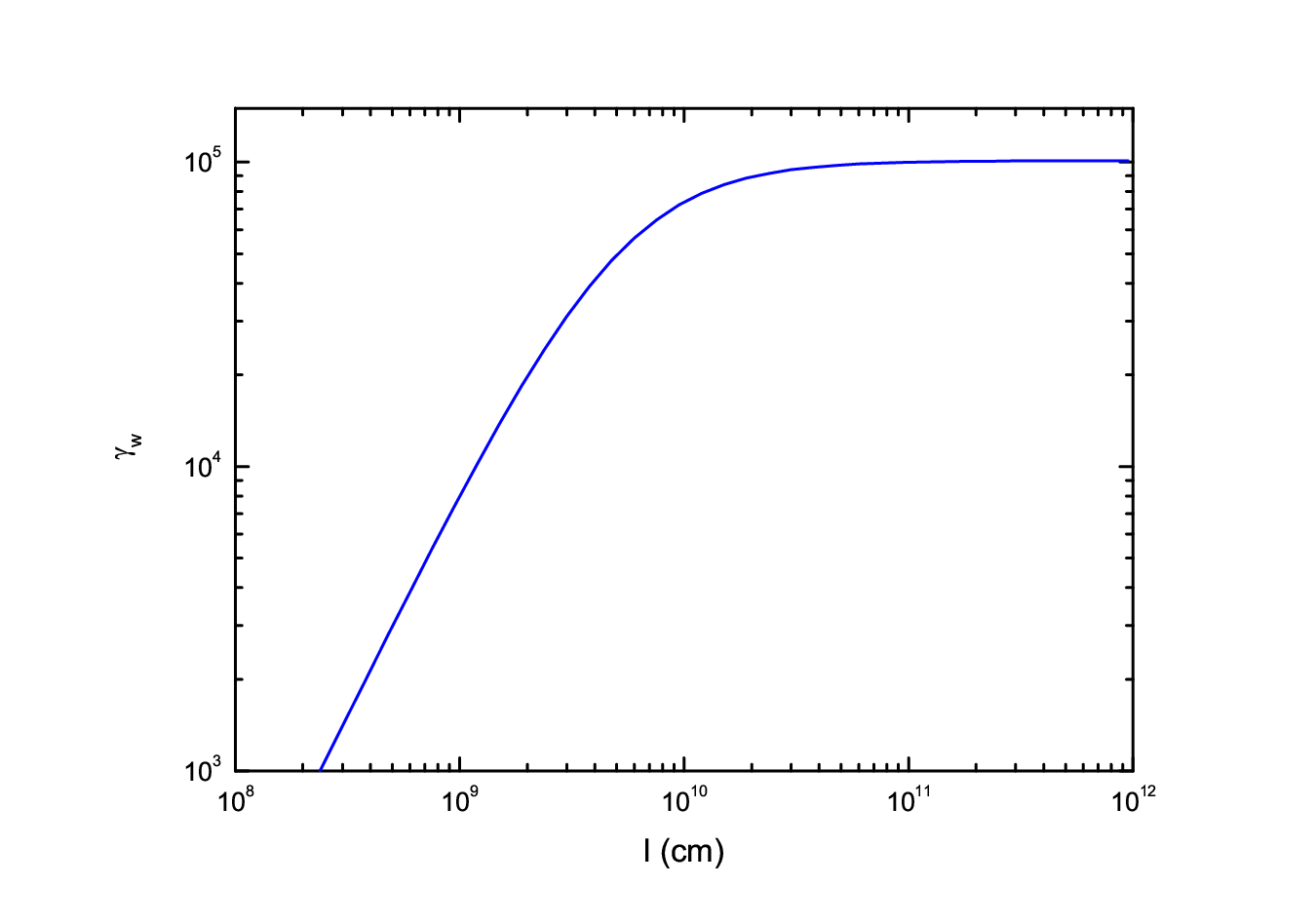}}
  \centerline{(a)}
\end{minipage}
\hfill
\begin{minipage}{0.45\linewidth}
  \centerline{\includegraphics[width=8.0cm]{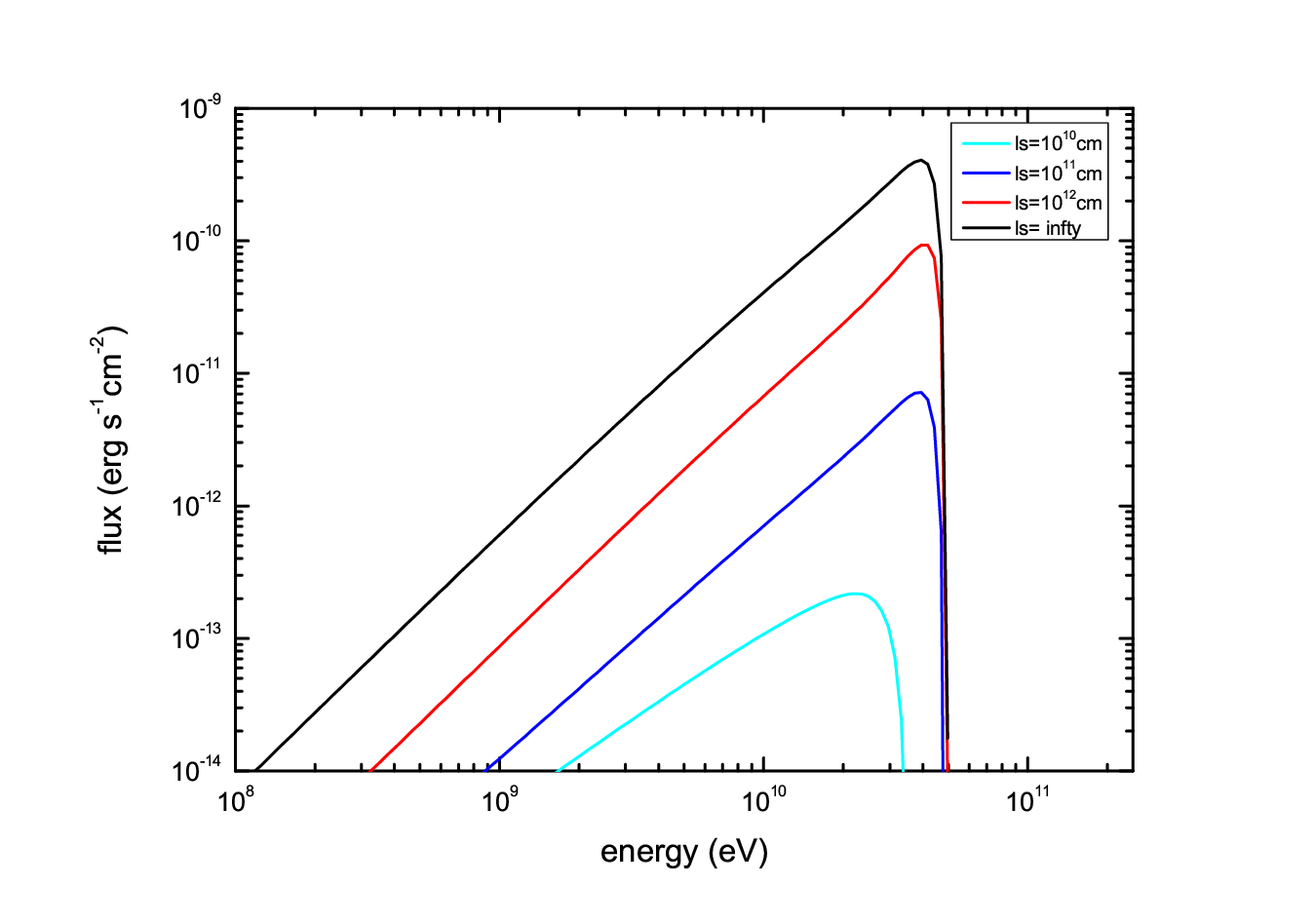}}
  \centerline{(b)}
\end{minipage}
\caption{Left: The radial evolution of pulsar wind LF; Right: The IC spectra with different travel distance of PWZ along LOS at SUPC. The model parameters adopted in calculations are $\sigma_{\rm L}=1\times10^2, \gamma_{\rm L}=1\times10^3$ and $\alpha_{\sigma}=1.5$.}
\label{fig:SED_PWZ}
\end{figure}

\begin{figure}
\begin{minipage}{0.45\linewidth}
  \centerline{\includegraphics[width=8.0cm]{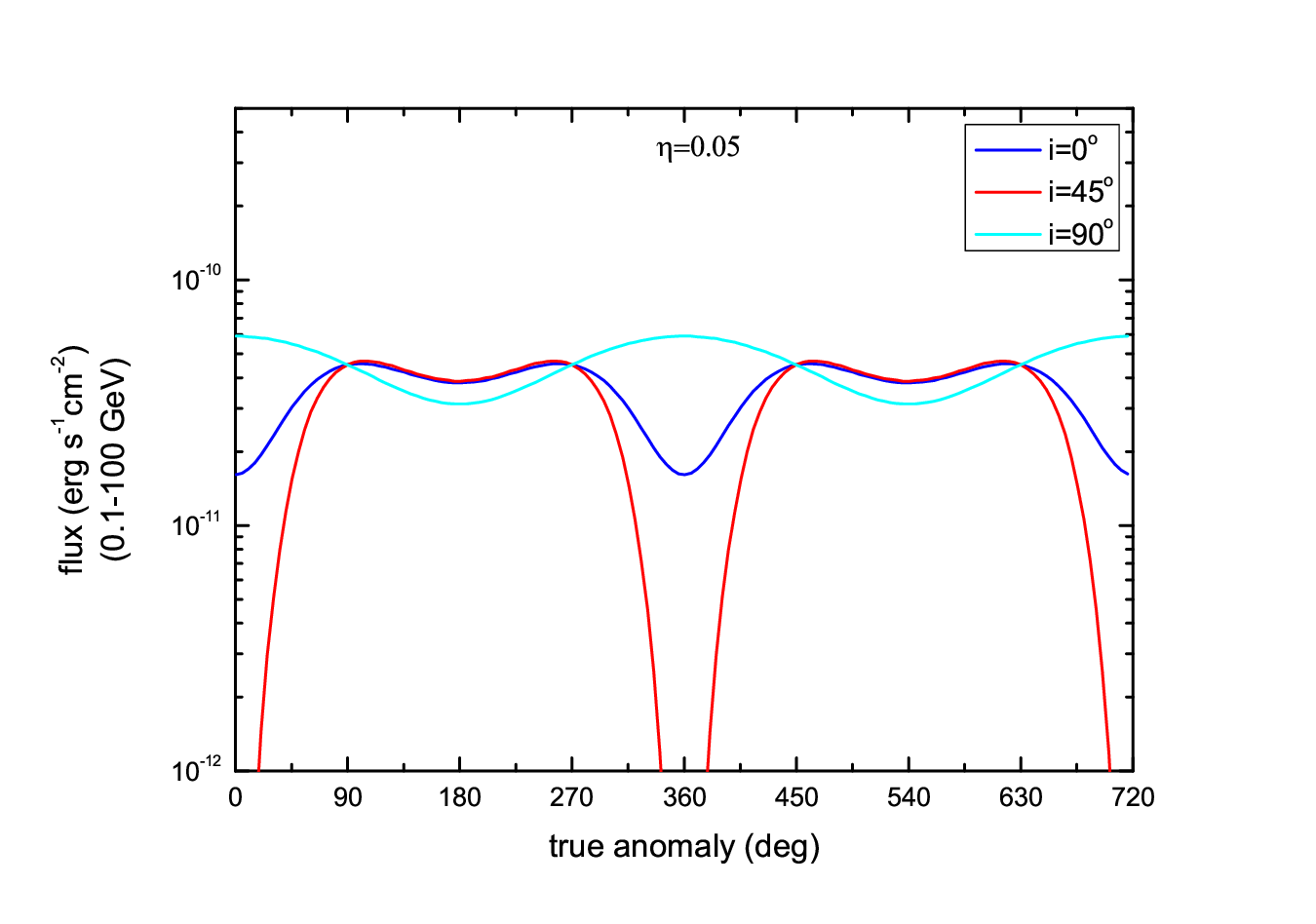}}
  \centerline{(a)}
\end{minipage}
\hfill
\begin{minipage}{0.45\linewidth}
  \centerline{\includegraphics[width=8.0cm]{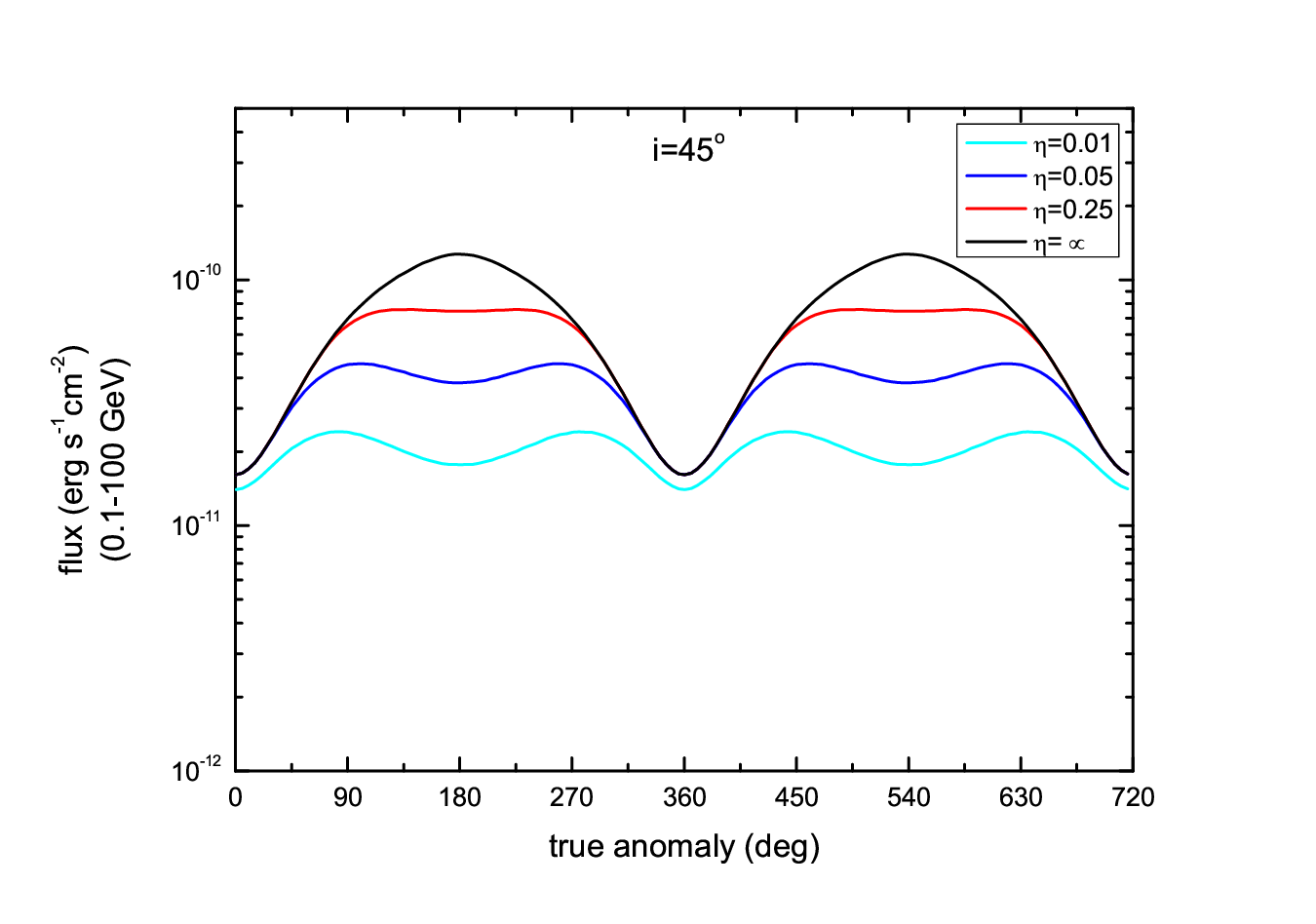}}
  \centerline{(b)}
\end{minipage}
\caption{The orbital modulations of HE $\gamma$-rays (0.1-100 GeV) from PWZ with different values of inclination angles ($i$) and momentum flux ratios ($\eta$).}
\label{fig:LC_PWZ}
\end{figure}

In the left panel of Fig.\ \ref{fig:SED_PWZ}, we present the radial evolution of pulsar wind LF. Initially, the pulsar wind is dominated by Poynting flux, and therefore the LF of pulsar wind particles at the light cylinder cannot be very large. With the dissipation of Poynting flux, the magnetic energy of pulsar wind would be gradually converted into kinetic energy of particles. As the magnetization parameter drops below unity, the LF of pulsar wind particles reaches its maximum with the order of $\gamma_{\rm w}\sim\sigma_{\rm L}\gamma_{\rm L}$.
The corresponding IC spectra with different travel distances of PWZ along LOS at SUPC are provided in the right panel of Fig.\ \ref{fig:SED_PWZ}.

In Fig. \ref{fig:LC_PWZ}, we present the orbital modulations of HE $\gamma$-ray flux in \textit{Fermi}/LAT energy band due to the IC scattering in PWZ with different values of inclination angles (left panel) and momentum flux ratios (right panel).
For a fixed shock structure (i.e., $\eta=0.05$), a larger inclination angle of the orbit $i$ predicts a more significant modulation of $\gamma$-ray flux due to the variation of the scattering angle. As the inclination angle is small enough (i.e., $i=0^{\circ}$), the $\gamma$-ray flux is mainly determined by the density of stellar photons, which shows flux maximum at periastron ($\phi=0^{\circ}$).
For a fixed viewing angle (i.e., $i=45^{\circ}$), a larger value of momentum flux ratio $\eta$ means that the size of the un-shocked wind is larger, and therefore the integrated $\gamma$-ray flux is higher.
When the pulsar is moving around INFC, the flux reaches its minimum due to the inefficient tail-on collision. We note that the light curves also feature some dips when the pulsar is moving around superior conjunction (SUPC, $\phi=\omega+180^{\circ}$), which is caused by the decrease of the size in PWZ along LOS and a larger binary separation. Alternatively, for a larger eccentricity, the dip would become more obvious.
For comparison, we also present the case of free expanding pulsar wind without the termination shock (i.e., $\eta=\infty$). As expected, the presence of the shock reduces the size of PWZ, and thus reduces the $\gamma$-ray flux especially when the pulsar is near SUPC.

\subsection{Emission model for the termination shock}\label{sect:TS}
As the pulsar wind is terminated by stellar outflows, the kinetic energy of particles would be converted into internal energy of the shock. According to magnetohydrodynamic shock jump conditions, the magnetic field at the shock can be obtained with
\begin{equation}\label{B}
B=\sqrt{\frac{L_{\rm sd}\sigma}{l_{\rm s}^2c(1+\sigma)}\left(1+\frac{1}{u^2}\right)},
\end{equation}
\begin{equation}\label{u}
u^2=
\frac{8\sigma^2+10\sigma+1}{16(\sigma+1)}+
\frac{[64\sigma^2(\sigma+1)^2+20\sigma(\sigma+1)+1]^{1/2}}{16(\sigma+1)},
\end{equation}
where $u$ and $\sigma$ are the radial four velocity and the magnetization of the wind, respectively (\citealt{Kennel1984a,Kennel1984b}).
Besides the compression of the magnetic field, the shock will also accelerate electron pairs into a power-law distribution $Q(\gamma)\propto \gamma^{-p}$. The accelerated electrons in the shock lose energies through radiative cooling and adiabatic process. The cooled spectrum of shocked electrons is given by (\citealt{Zabalza2013,Chen2019})
\begin{equation}\label{ne}
n(\gamma)=\frac{1}{|\dot{\gamma}|}
\int Q(\gamma^{\prime}) {\rm d} \gamma^{\prime},
\end{equation}
with $\dot{\gamma}$ being the total energy loss rate (\citealt{Moderski2005}; \citealt{Khangulyan2007}). Then the local emissivity of the shock can be calculated by:
\begin{equation}\label{j}
  j(\nu)=\int n(\gamma)P(\gamma)\rm{d}\gamma,
\end{equation}
where $P(\gamma)$ is the total power of synchrotron and IC scattering for a single electron (\citealt{Kirk1999}).
The synchrotron power is given by (\citealt{Rybicki1979})
\begin{eqnarray}\label{PSYN}
P(\gamma) &=& \frac{\sqrt{3}q_{\rm e}^3B}{m_{\rm e}c^2}F\left(\frac{\nu}{\nu_{\rm c}}\right), \\
  \nonumber F(x) &=& x\int_x^{\infty}K_{5/3}(y){\rm d}y,
\end{eqnarray}
where $\nu_{\rm c}=3\gamma^2q_{\rm e}B/4\pi m_{\rm e}c$ is the characteristic frequency and $K_{5/3}$ is the modified Bessel function. As for IC scattering, we only consider the external IC emission with seed photons from the massive star, and synchrotron-self-Compton emission is ignored due to strong suppression of the Klein-Nishina effect (\citealt{Dubus2006b,Kong2011}).
The IC scattering power for a single electron is given in Eq. (\ref{PEIC}).

According to numerical simulations of gamma-ray binaries, the shocked flow would propagate in a narrow region with an increasing bulk velocity (\citealt{Bogovalov2008,Bogovalov2012}). It means that the shock emission from the tail could be highly beamed. In particular, as the beaming direction passes through LOS, we will receive the boosted emission from the shock tail (\citealt{Dubus2010,Dubus2015}). Taking the Doppler-boosting effect into consideration, the total flux from the bow shock is given by (e.g., \citealt{Granot1999}):
\begin{equation}\label{F}
  F(\nu)=\frac{1}{d_{\rm L}^2}\int_{\theta_{\rm sh}}^{\pi}\sin\theta{\rm{d}}\theta \int_0^{2\pi}{\rm{d}}\varphi \int_{l_{\rm s}}^{l_{\rm s}+\Delta_{\rm s}}r^2{\rm{d}}r D^2j(\nu/D)\exp(-\tau),
\end{equation}
where $\Delta_{\rm s}\sim 0.1 l_{\rm s}$ is the shock thickness, and $\theta$ and $\varphi$ are the polar and azimuthal angles, respectively, of the flow measure from the symmetric axis of the shock cone.
The Doppler factor is
\begin{equation}\label{D}
  D=\frac{1}{\Gamma(1-\beta\cos\alpha)},
\end{equation}
with $\Gamma$ being the bulk LF of the moving flow elements, and $\beta=\sqrt{1-\Gamma^{-2}}$. For simplicity, we assume that the shocked flows are moving with the same speed. The angle between the shocked flow and LOS is given by (e.g., \citealt{Kathirgamaraju2018})
\begin{equation}\label{vartheta}
  \cos\alpha=\cos\theta_{\rm o}\cos\theta_{\rm s}+\sin\theta_{\rm o}\sin\theta_{\rm s}\cos\varphi,
\end{equation}
where $\theta_{\rm o}=\pi-\psi_0$ is the angle between the symmetric axis of the shock cone directed radially away from the star and LOS. For a purely radial shock, maximum boosting happens around INFC where the flow elements are moving towards us.

Apart from the boosting effect, the VHE $\gamma$-rays would be absorbed by the soft stellar photons. In particular, the absorption could be important when the pulsar is moving around SUPC due to the huge amounts of soft photons along LOS. The optical depth due to pair creation can be calculated as
\begin{equation}\label{tau}
  \tau=\int {\rm d}l\int {\rm d}\nu_{\rm s}(1-\mu)n_{\rm ph}(\nu_{\rm s})\sigma_{\gamma\gamma},
\end{equation}
where $n_{\rm ph}$ is the number density of stellar photons, $\mu=\cos\theta_{\gamma\gamma}$ is the collision angle and $\sigma_{\gamma\gamma}$ is the cross-section of pair creation (\citealt{Gould1967}). A detailed description of $\gamma$-ray absorption in binaries can be found in \cite{Dubus2006b}.

\begin{figure}
\begin{minipage}{0.45\linewidth}
  \centerline{\includegraphics[width=8.0cm]{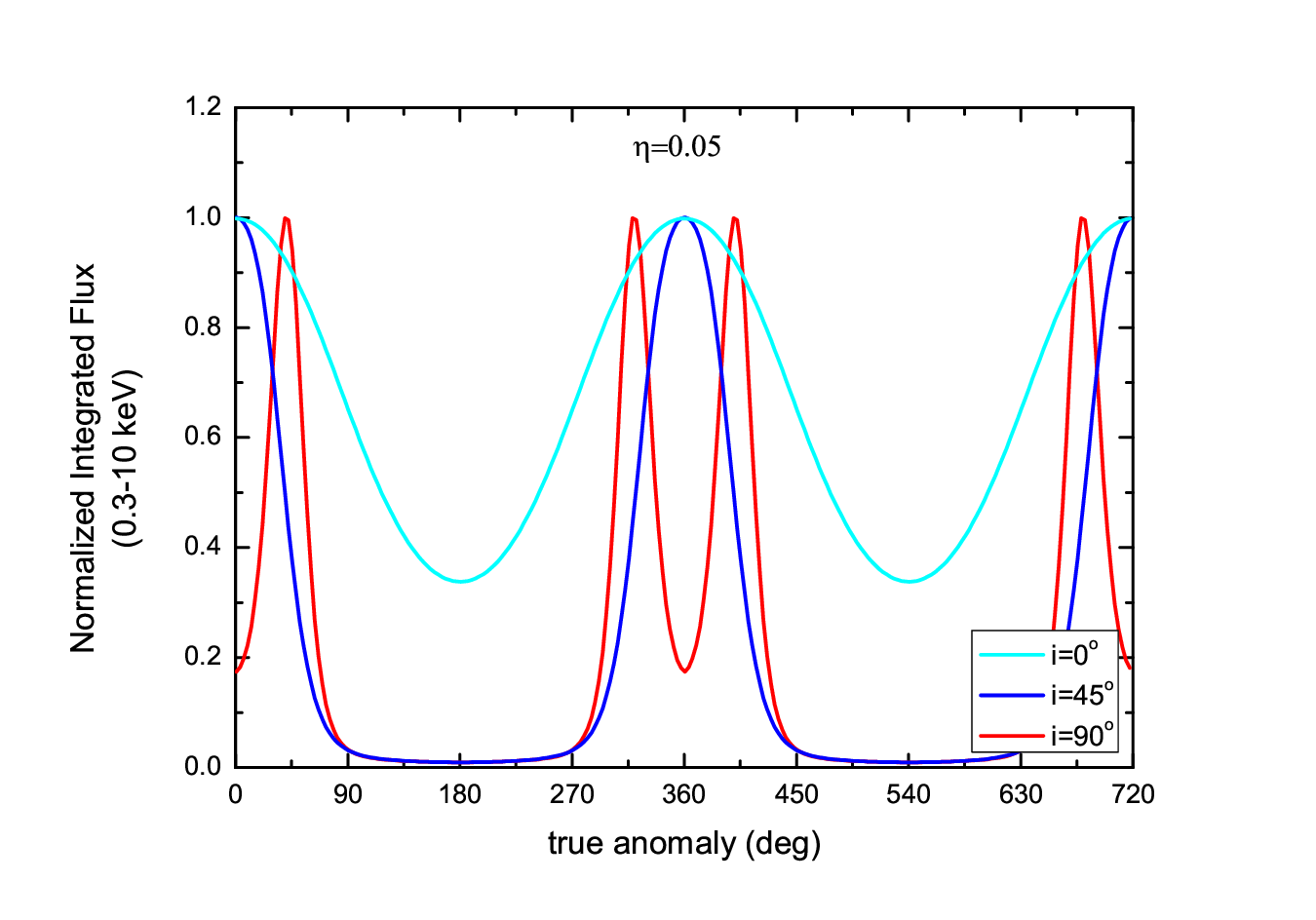}}
  \centerline{(a)}
\end{minipage}
\hfill
\begin{minipage}{0.45\linewidth}
  \centerline{\includegraphics[width=8.0cm]{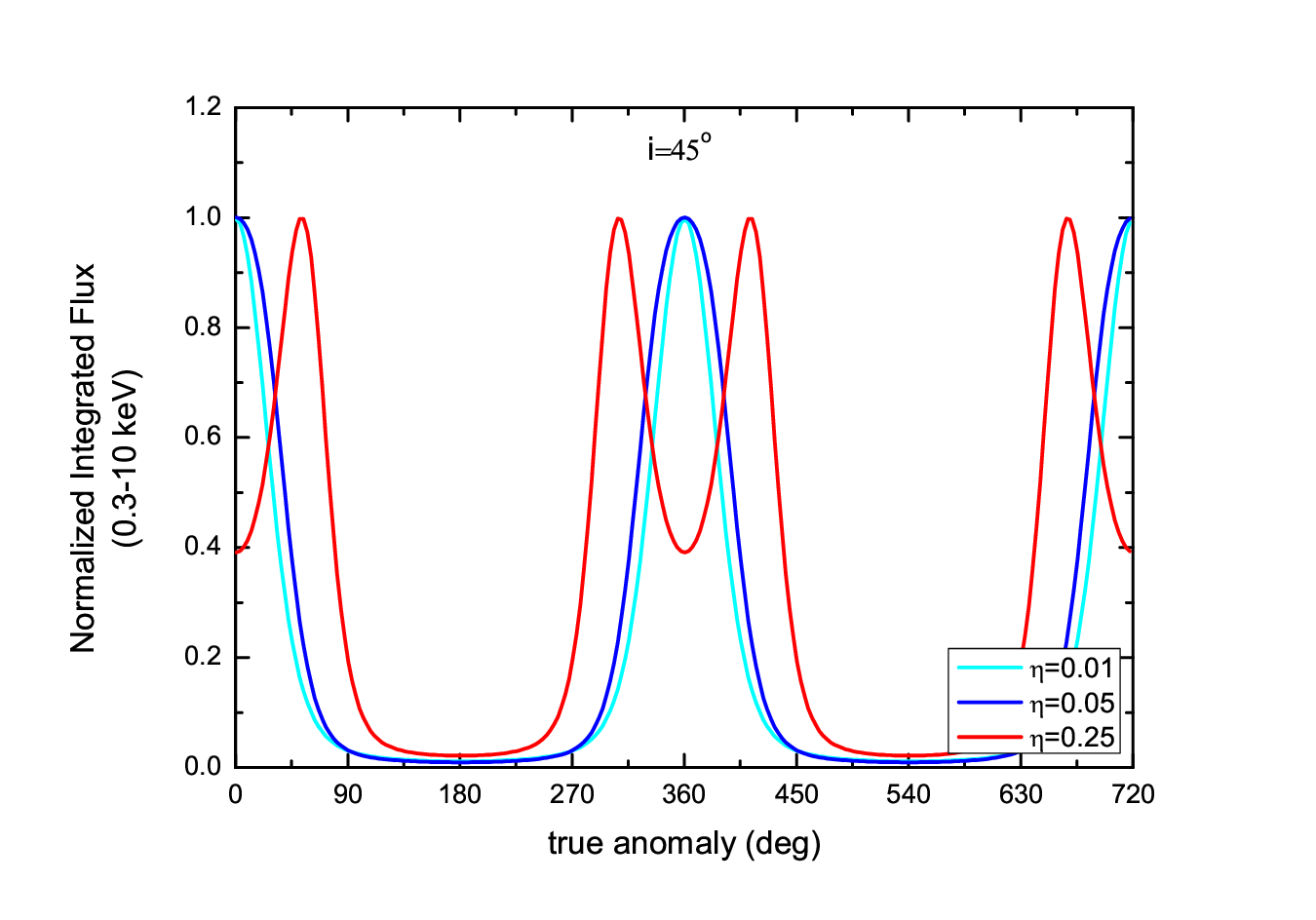}}
  \centerline{(b)}
\end{minipage}
\vfill
\begin{minipage}{0.45\linewidth}
  \centerline{\includegraphics[width=8.0cm]{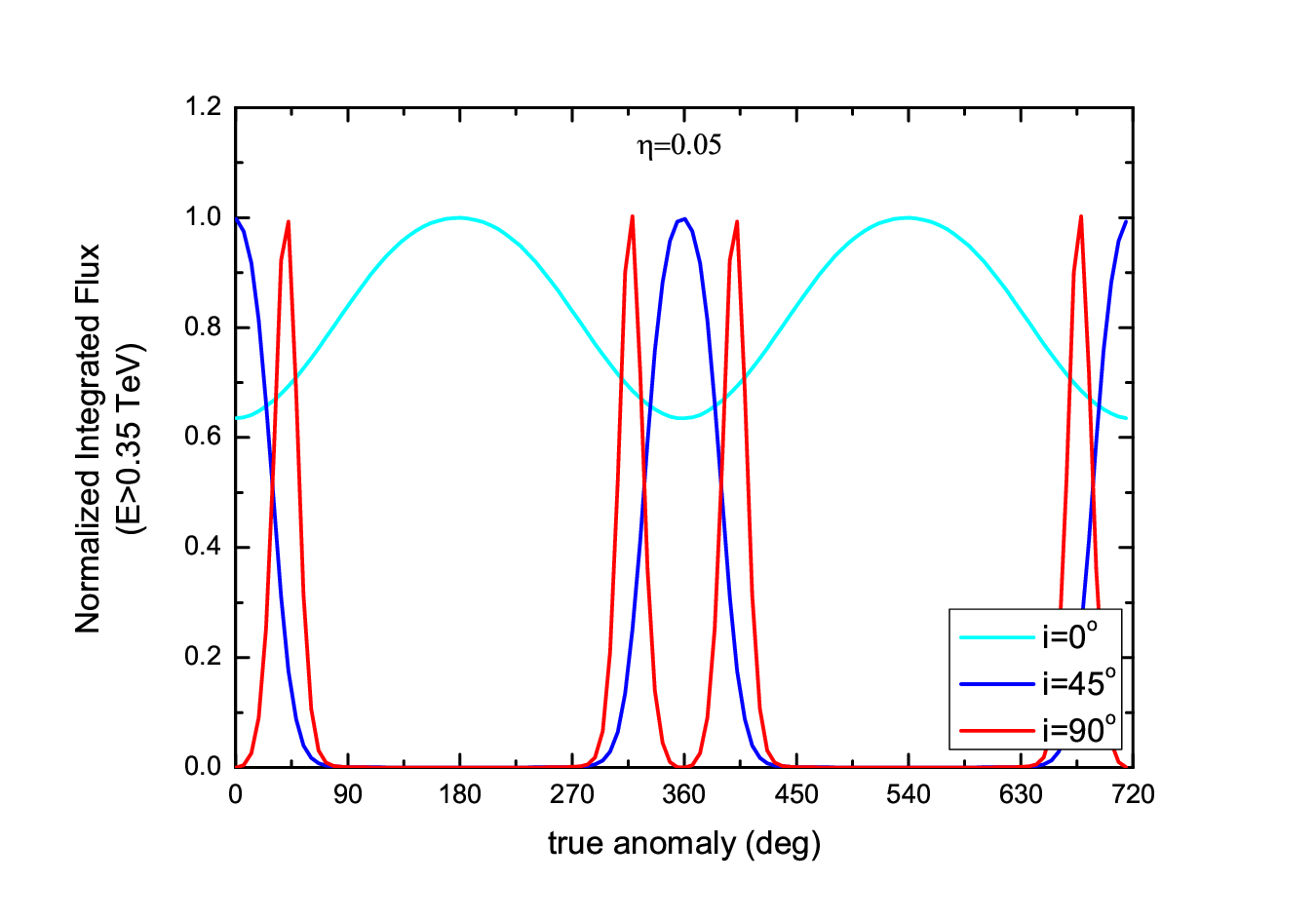}}
  \centerline{(c)}
\end{minipage}
\hfill
\begin{minipage}{0.45\linewidth}
  \centerline{\includegraphics[width=8.0cm]{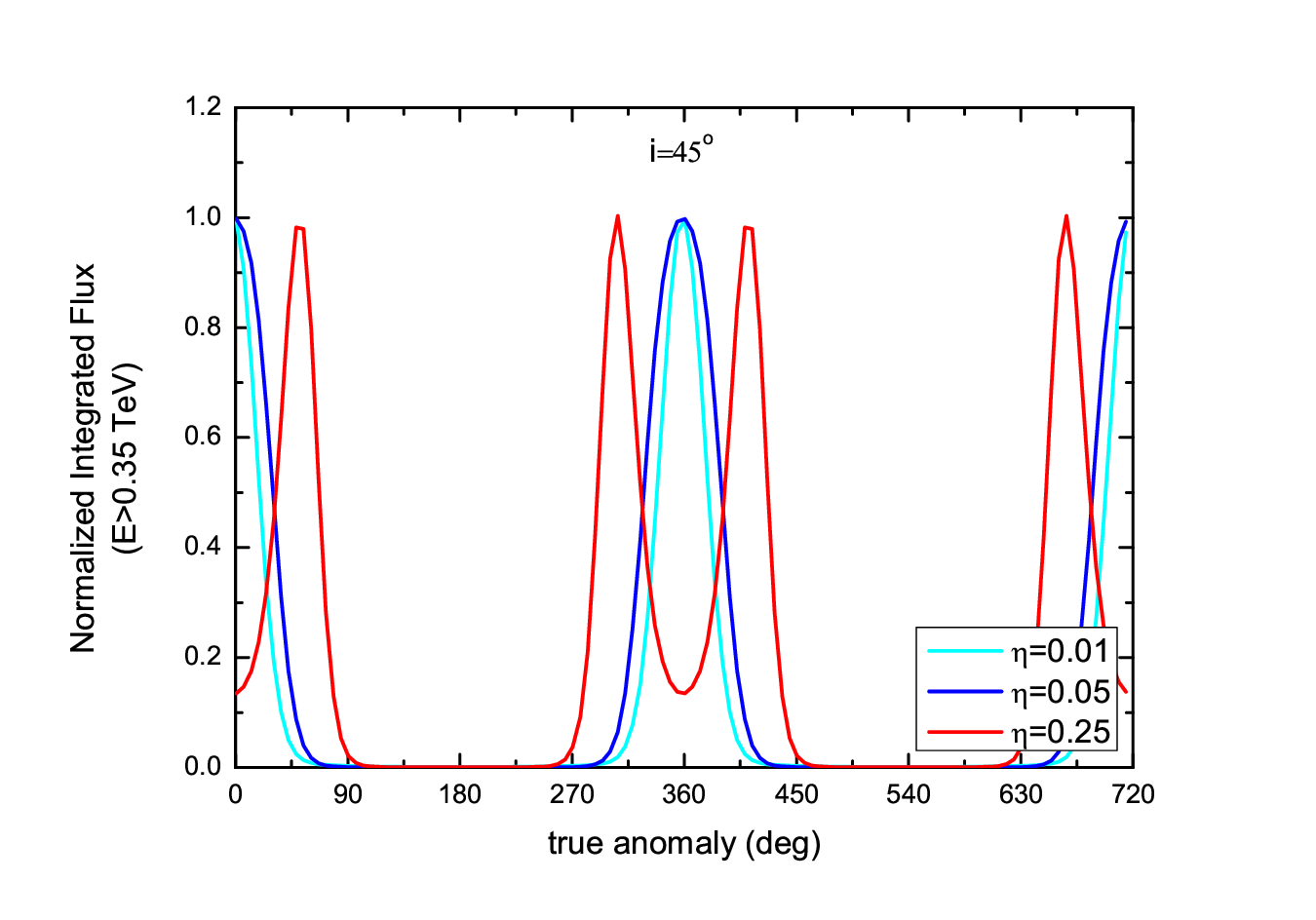}}
  \centerline{(d)}
\end{minipage}
\caption{The normalized integrated fluxes in X-ray (0.3-10 keV, upper panels) and VHE $\gamma$-ray  ($E>0.35$ TeV, bottom panels) from the termination shock with different values of momentum flux ratio and orbital inclination angles.}
\label{fig:LC_IBS}
\end{figure}

Among all model parameters, the orbital inclination angle and the momentum flux ratio between two winds are the most uncertain ones, which would significantly affect the orbital modulations of shock radiations. Therefore, it is necessary to explore the effects of the parameters on light curves.
In Fig. \ref{fig:LC_IBS}, we show the normalized integrated flux from IBS with the Doppler-boosting effect in X-ray (0.3-10 keV, upper panels) and VHE $\gamma$-ray ($E\geq0.35\ $TeV, bottom panels). Since the shock geometry is determined by the momentum flux ratio, and the orbital motion of the pulsar around the companion leads to the rotation of the shock cone, the shock radiation received by the observer changes significantly with the angle between LOS and the moving direction of the shocked flow. Depending on the relations between the inclination angle of the orbit $i$ and the opening angle of the shock $\theta_{\rm sh}$ (which is governed by the momentum flux ratio $\eta$), two different patterns of light curves will be observed.
In particular, when the inclination angle satisfies $\pi/2-i<\theta_{\rm sh}$, LOS will pass through the shock cone twice every orbit, and two rapid flares will be observed around INFC due to the boosted emission from the shock region. The positions of the two flux maxima occur when the shock is passing through LOS, which are given by:
\begin{equation}
  \phi_{1,2}=\phi_{\rm{INFC}}\pm\Delta\phi,
\end{equation}
where $\phi_{\rm{INFC}}=\omega$ is the anomaly of INFC, and
\begin{equation}
  \Delta\phi=\arccos\left(\frac{\cos\theta_{\rm sh}}{\sin i}\right).
\end{equation}
The duration of each flare can be approximated with $2/\Gamma\times180^{\circ}/\pi$ of orbital phase, and a larger value of $\Gamma$ would definitely lead to sharper flares around INFC. As the inclination angle of the orbit decreases to $i\leq\pi/2-\theta_{\rm sh}$ or the opening angle of the shock satisfies $\theta_{\rm sh}\leq\pi/2-i$ (i.e., a smaller value of $\eta$), two sharp peaks at $\phi_1$ and $\phi_2$ would be merged at INFC and finally disappear (\citealt{Neronov2008}).

For the case of $i=0$ (i.e., LOS is perpendicular to the orbital plane), the Doppler factor is constant throughout the orbit. Since the X-rays are generated by synchrotron radiation, the X-ray intensity mainly depends on the magnetic field strength in the shock. Under the assumption that the magnetization of un-shocked wind evolves with radial distance in the form of $\sigma\propto l^{-\alpha}$, the magnetic field strength in the termination shock would be higher as the shock is closer to the pulsar. So, the X-ray light curves display flux maximum at periastron.
As for TeV emission produced by IC scattering, the flux modulations are much more complicated due to a combination of effects, including the orbital variations of scattering angles between relativistic electrons and soft photons, and the pair creation process. Although the stellar photon density achieves its highest value at periastron, the gamma-ray absorption and the rapid cooling process of electrons will further reduce the TeV flux, and therefore features a dip at periastron.
We should note that in the calculations of Fig. \ref{fig:LC_IBS}, we assume that the rotating hollow cone has a symmetric axis directed radially away from the star. In particular, when the periastron is assumed to be at INFC (which is the case for J1018), the observed light curves from the shocked cone have a symmetric profile around INFC. When the periastron passage of the pulsar is not at the INFC, then the light curve profiles are not symmetric due to the orbital modulations of the magnetic field and the photon field.

\section{Fitting Results}\label{sect:results}
In this section, we utilize the emission model described above to calculate emissions from J1018.
We use the orbital solution as obtained by \cite{Monageng2017}, with the eccentricity of $e=0.31$, and the periastron phase occurring around INFC ($\omega\sim1^{\circ}$).
The LF and magnetization parameter at the light cylinder are taken as $\gamma_{\rm L}=10^3$ and $\sigma_{\rm L}=10^2$, respectively, with the decay index of the magnetization parameter of $\alpha_{\sigma}=1.5$ (\citealt{Kong2011,Takata2017}).
The photon index in 3-10 keV during the entire orbit is $\Gamma_{\rm X}\simeq1.2-1.8$ which suggests that the power-law index of injected electrons in the shock is in the range of $p=2\Gamma_{\rm X}-1\simeq1.4-2.6$. Given the flat spectrum at VHE $\gamma$-ray band, we adopt $p=2.1$ in calculations.
The shocked flow is assumed to move with a mildly-relativistic speed with a bulk LF of
$\Gamma=2$ as adopted in \cite{Kong2012}. The other parameters of the pulsar are obtained by fitting the steady component detected by \textit{Fermi}/LAT as we discuss below. The related model parameters for J1018 are listed in Table. \ref{table}.

\subsection{Outer gap emission from J1018}
\begin{figure}
\centerline{
\includegraphics[width=12cm,height=8cm]{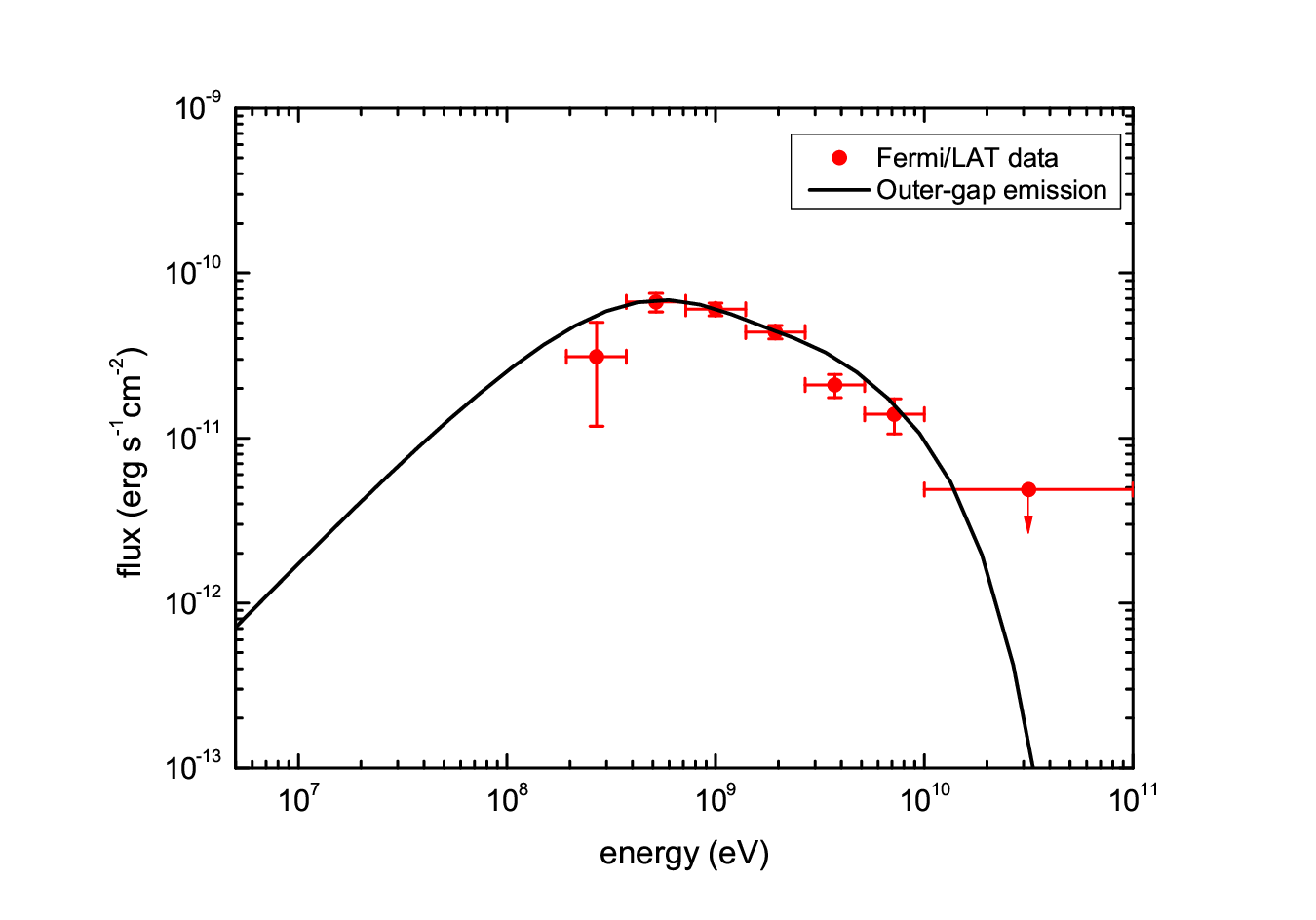}}
\caption{The calculated outer gap emission with a comparison of the steady component detected by \textit{Fermi}/LAT.}
\label{fig:outer_gap}
\end{figure}
Currently, because there is no existing result on the
timing parameters of the pulsar in J1018, the properties of the
pulsar remain unknown. To explain the complete
emission spectrum of J1018, the magnetospheric contribution cannot
be ignored. We use the standard outer gap model
to simulate the curvature spectrum from the outer gap, which extends from the null charge
surface to the light cylinder (\citealt{Cheng1986a,Cheng1986b}).
The separation of the oppositely
charged particles induces an electric potential in the space between
them, leading to the growth of the outer gap. On the other hand, the
curvature photons can undergo pair creation with the
softer photons from the pulsar surface. The accumulation of charges
will reduce the electric potential, resulting in depletion of the
outer gap. These two instantaneously occurring processes can be
approximately modeled by the two-layer structure, which defines contrasting charge densities for the primary acceleration and the screening regions.

In this study, we follow the two-layer outer gap model explored by \cite{Wang2010,Wang2011}, in which the accelerating electric field is solved by assuming the charge density in the outer gap. We assume a moderate value for
the charge density in the gap, namely, that the gap has a charge density of 70\% of the Goldreich-Julian value. We solve a two dimensional Poisson equation in the poloidal plane and assume that there is no variation of the gap structure in the azimuthal direction. This approximation will be justified if the thickness of
the gap in the poloidal plane is much smaller than the width of the azimuthal direction, for which
we apply $\sim 180^{\circ}$. The spin period and the surface dipole magnetic field strength are $0.05\ {\rm s}$ and $10^{12}$~G, respectively, which yield a spin-down power of $L_{\rm sd}\simeq 1\times 10^{36} {\rm erg~s^{-1}}$. By assuming the pulsar has an inclination angle of $40^{\circ}$, we calculate the direction of $\gamma$-rays
at each calculation point and obtain the spectrum as a function of the viewing angle. In this paper, we present the calculated $\gamma$-ray spectrum for the  viewing angle at $120^\circ$ (or $ 60^\circ)$, which provides a reasonable fit to the GeV spectrum in the LOW state as plotted in Fig. \ref{fig:outer_gap}.

\begin{figure}
\centerline{
\includegraphics[width=12cm,height=8cm]{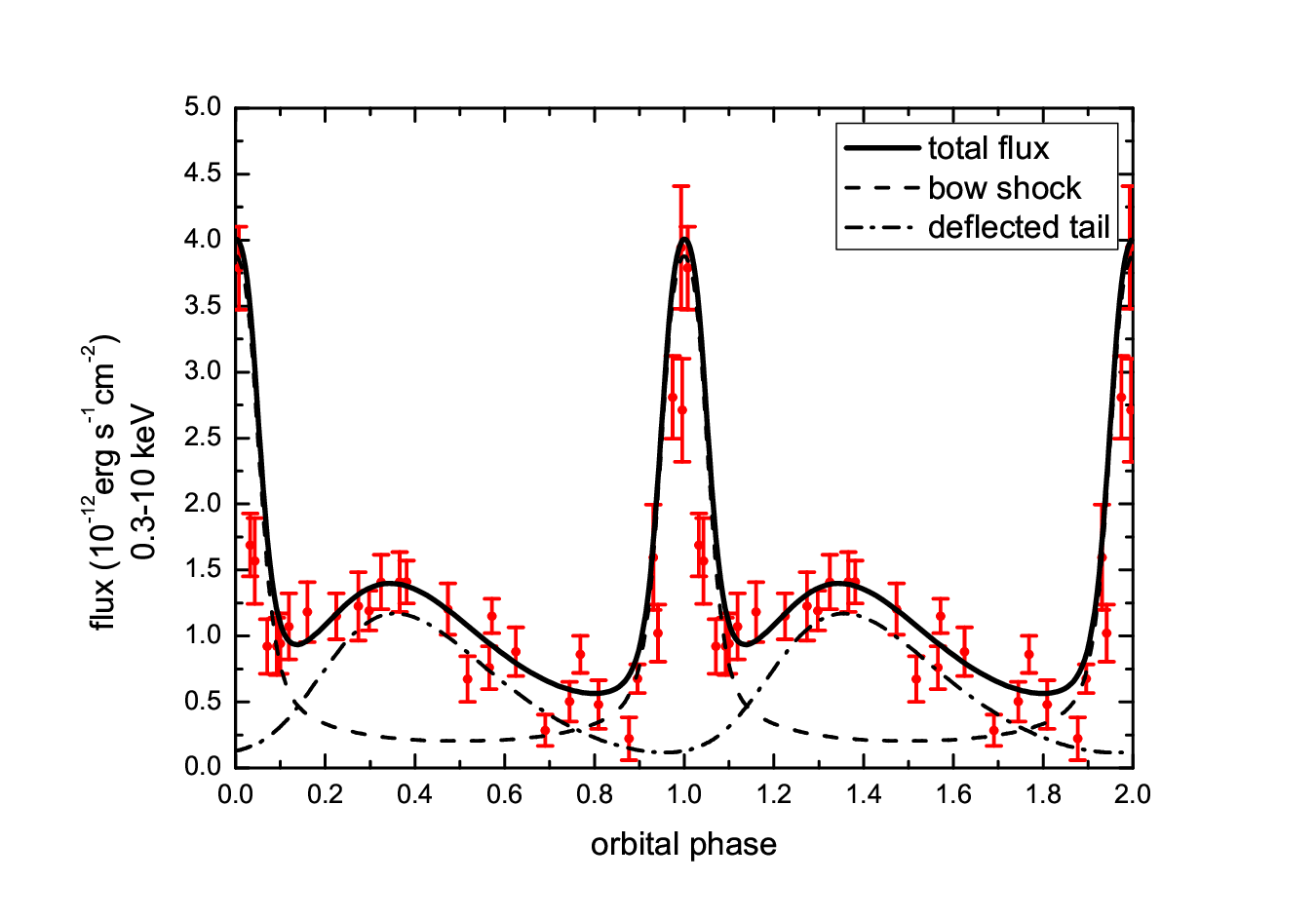}}
\caption{The integrated X-ray ($0.3-10$keV) flux of J1018 as compared with observational data.
The dashed line corresponds to the shock emission, while the dash dotted line signifies the contribution from the deflected tail. The X-ray data are taken from \citealt{Fermi2012}.} \label{fig:keV}
\end{figure}

\begin{figure}
\centerline{
\includegraphics[width=12cm,height=8cm]{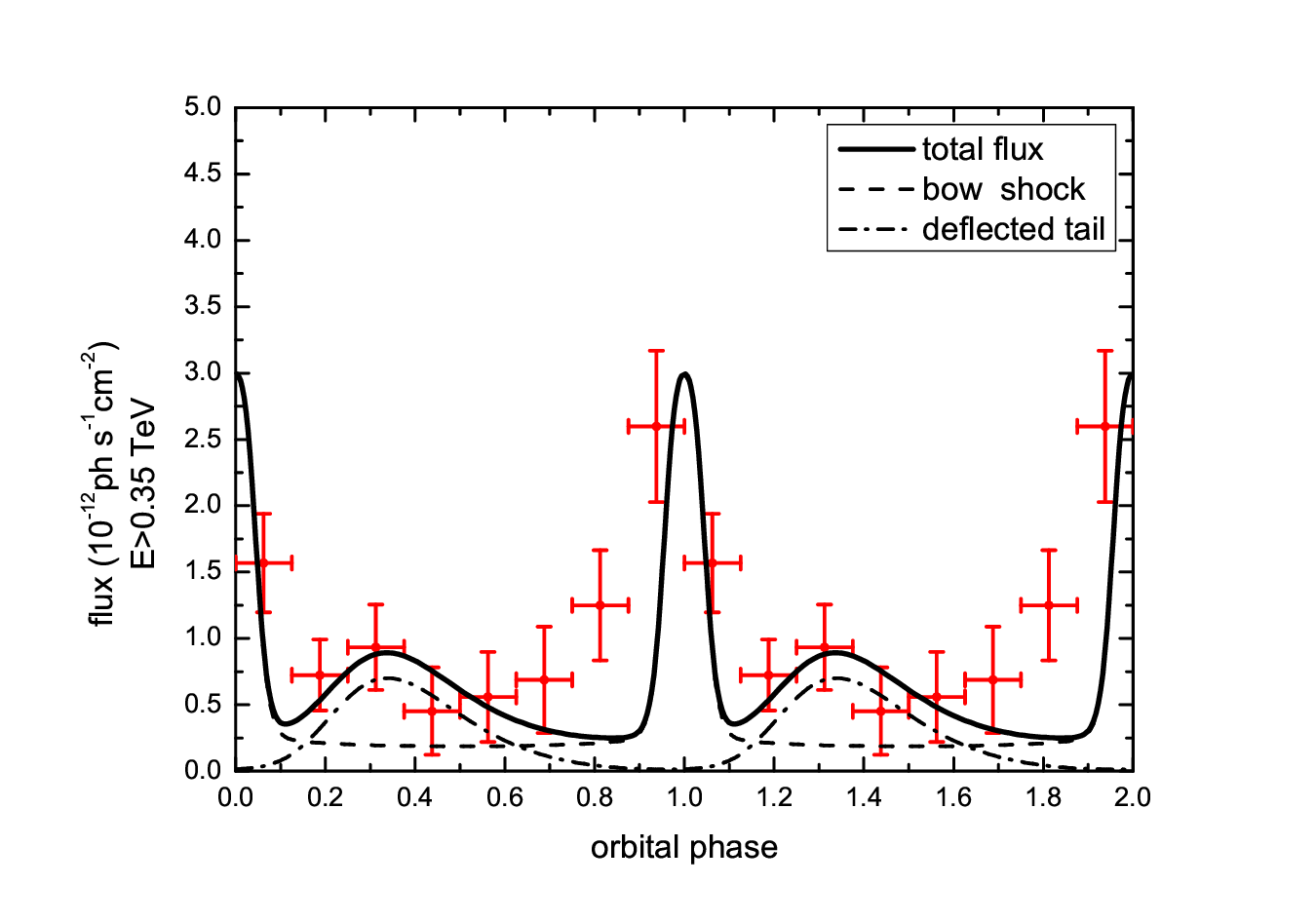}}
\caption{The integrated VHE $\gamma$-ray ($E>0.35$TeV) flux of J1018 as compared with observation data. The dashed line corresponds to the shock emission, while the dash dotted line signifies the contribution from the deflected tail. The $\gamma$-ray data are taken from \cite{HESS2015}. } \label{fig:TeV}
\end{figure}

\begin{figure}
\begin{minipage}{0.485\linewidth}
  \centerline{\includegraphics[width=8cm]{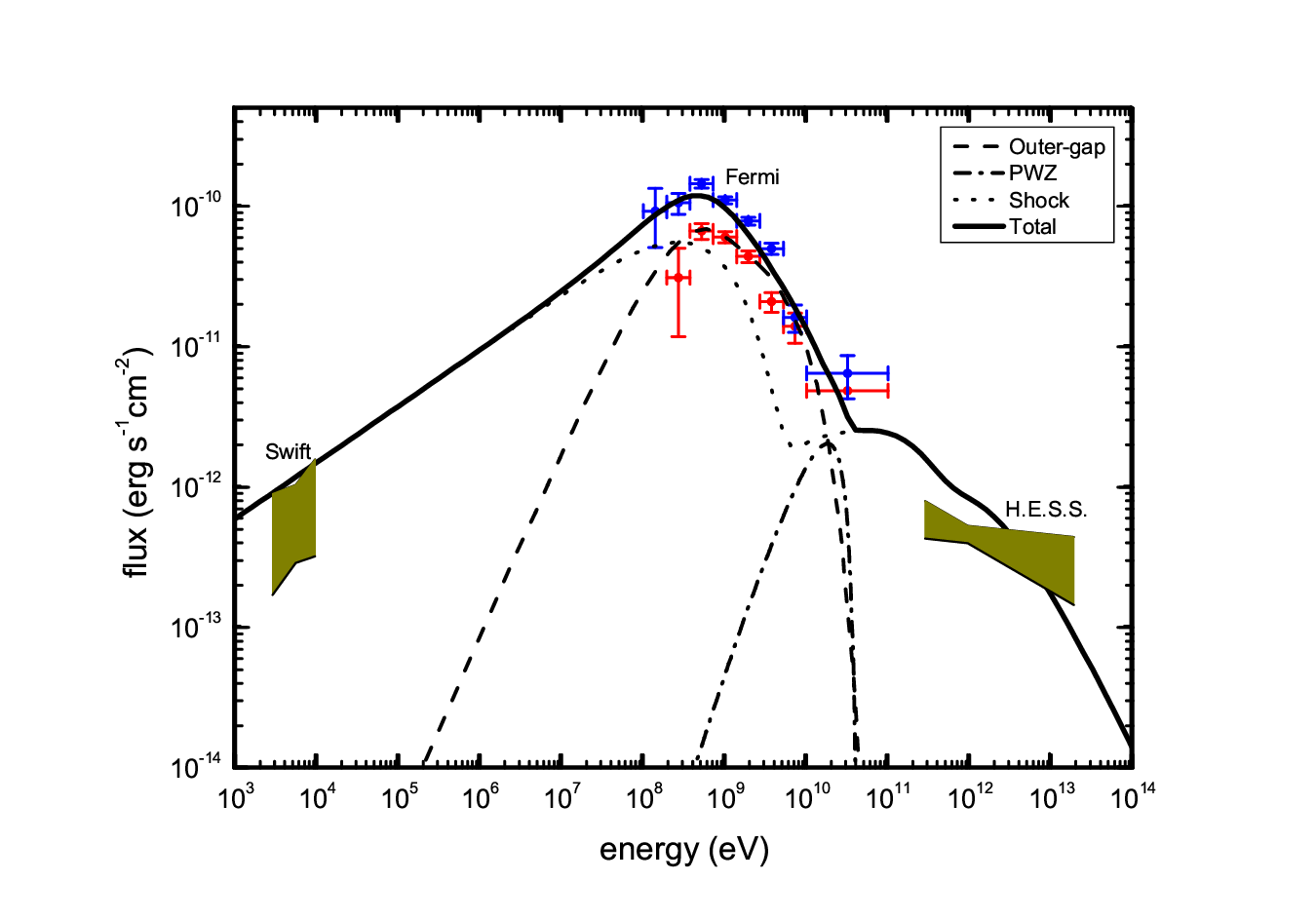}}
  \centerline{(a)}
\end{minipage}
\hfill
\begin{minipage}{0.485\linewidth}%
  \centerline{\includegraphics[width=8cm]{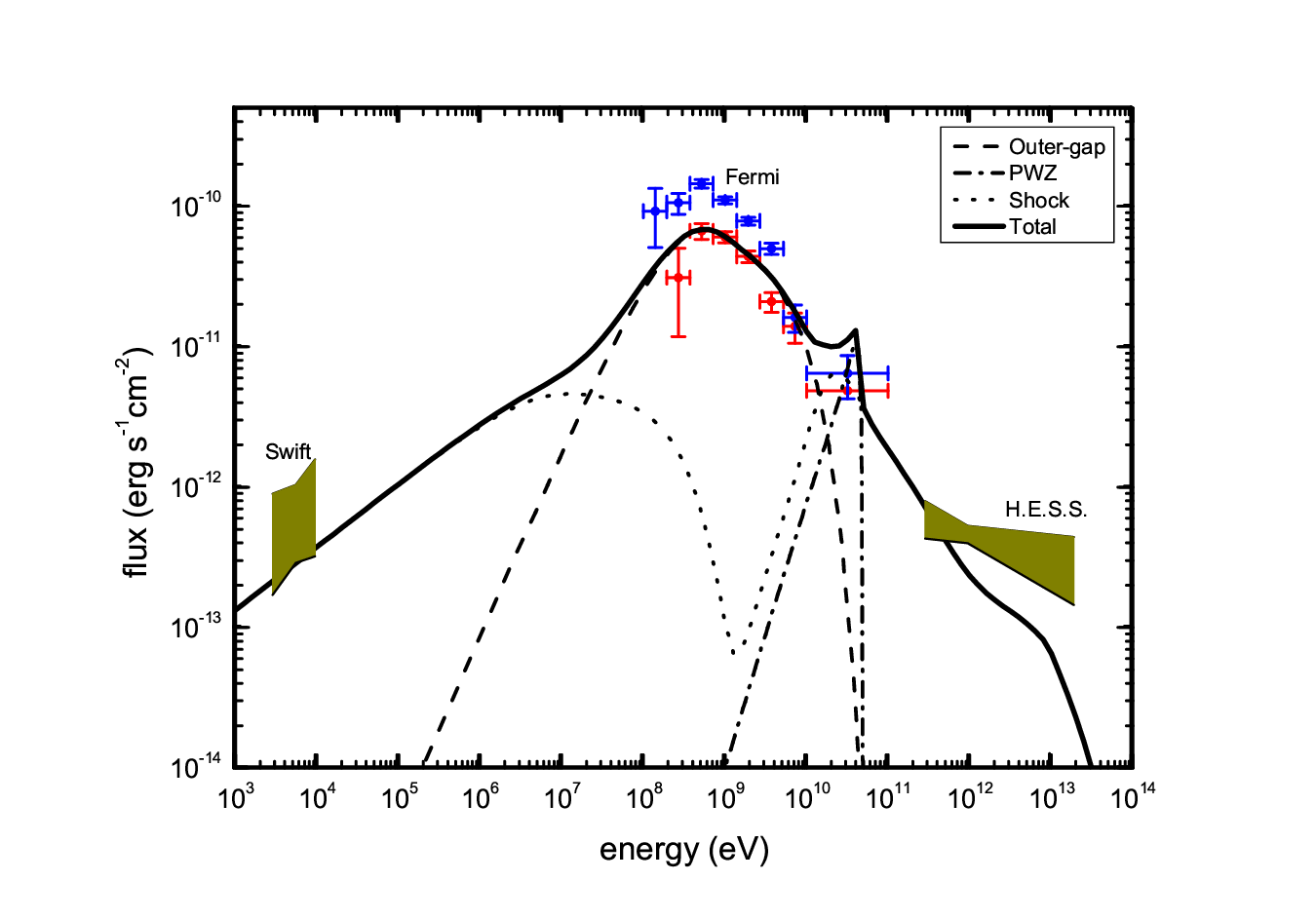}}
  \centerline{(b)}
\end{minipage}
\caption{Multi-wavelength spectra of J1018 at INFC (left panel) and SUPC (right panel). The \textit{Swift} and \textit{H.E.S.S.} data are taken from \cite{An2017} and \cite{HESS2015}, respectively. The solid lines are the model results including the emissions of pulsar magnetospheric (dashed lines), un-shocked wind (dash dotted lines) and termination shock (dotted lines).}
\label{fig:SED}
\end{figure}
\subsection{Modeling the HE Emissions from J1018}

\begin{figure}
\centerline{
\includegraphics[width=12cm,height=8cm]{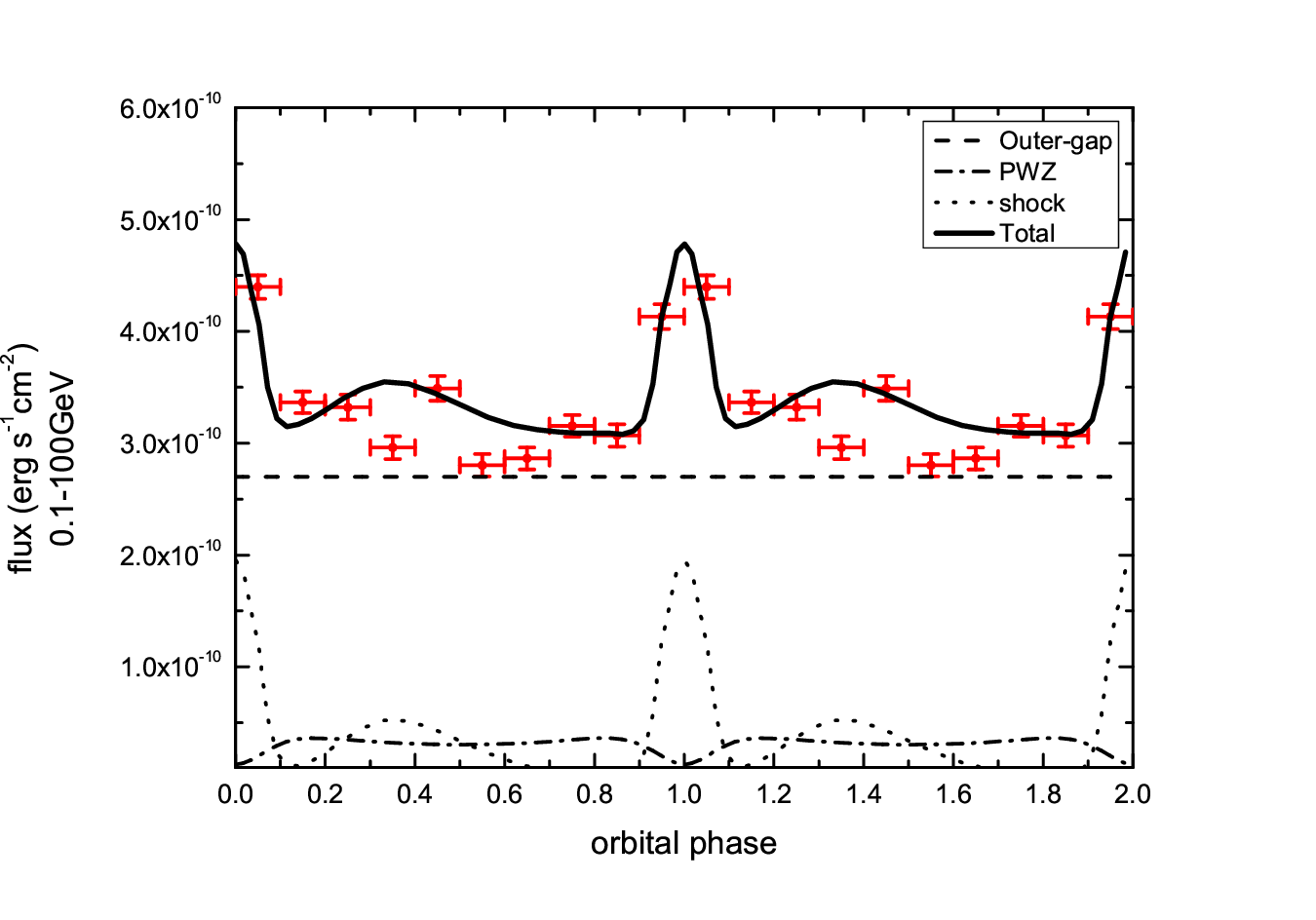}}
\caption{The orbital light curve of 0.1-100 GeV emission from J1018. The solid line corresponds to the model result including the emissions of pulsar magnetospheric (dashed line), un-shocked wind (dash dotted line) and termination shock (dotted line).}
\label{fig:GeV}
\end{figure}

In this subsection, we use the emission model described above and parameters listed in Table \ref{table} to fit the multi-wavelength emissions of J1018.
As shown in Fig. \ref{fig:keV} and \ref{fig:TeV}, the rapid flares around phase 0 in the X-ray and VHE $\gamma$-ray light curves of J1018 can be naturally explained by the boosted emissions from the shock as LOS passes around the shock cone at INFC. With a modest value of inclination angle $i\sim45^{\circ}$, the half opening angle of the shock should be less than $90^{\circ}-i$, otherwise, two sharp spikes will be observed in one orbital period, and therefore the momentum flux ratio of J1018 is expected to be less than $\eta\leq0.07$. Considering typical values of mass-loss rate and wind velocity of type O stars with $\dot{M}\sim10^{-7}M_{\odot}\cdot{\rm yr^{-1}}$ and $v_{\rm w}\sim10^{8} {\rm cm\cdot s^{-1}}$ respectively, the spin-down power is expected to be less than $L_{\rm sd}\leq1.5\times10^{36} {\rm erg\cdot s^{-1}}$.

Besides the rapid flare around phase 0, the X-ray light curve also exhibits a broad sinusoidal modulation component which peaks around phase $0.3-0.4$. We note that the radio light curve of J1018 also displays a smooth sine-wave modulation with flux maxima around phase $0.2-0.4$ (\citealt{Fermi2012}). This indicates that the broad sinusoidal modulations in X-ray and radio bands may have a common origin. The radio emission is believed to be produced by the tail of shocked flows at a larger distance (\citealt{Takata2009}), so we expect that this sinusoidal component could be also caused by the shock tail.
According to the hydrodynamical simulation, the Coriolis force due to fast orbital motion of the pulsar could amplify the bending of shocked flow (\citealt{Bosch-Ramon2012,Bosch-Ramon2015}), and it means that the flow direction at a larger distance of the tail is not radial. The realistic geometry of the shock tail could be very complicated. In our calculation, we simply treat the shock tail as a comet-like geometry starting at the distance of $l\geq 3d$ with a deflection angle of $\theta_{\rm flow}\simeq 150^{\circ}$.
In this case, the angle $\cos\theta_{\rm obs}$ in Eq.(\ref{vartheta}) should be replace with $\cos\theta_{\rm obs}=\sin i\cos(\phi-\omega-\theta_{\rm flow})$. The deflection of the shock tail at larger distance explains why the peak phase of this sinusoidal modulation is not around INFC (\citealt{Dubus2010}). Following the above description, we calculate the X-ray and VHE $\gamma$-ray emissions from the deflected tail as presented by dash dotted lines in Fig. \ref{fig:keV} and \ref{fig:TeV}, respectively.

The multi-wavelength spectra of J1018 with comparisons of the observational data at INFC (left panel) and SUPC (right panel) are presented in Fig.\ref{fig:SED}. The solid line is the total flux from the binary system, including the curvature emission from the outer gap (dashed line), IC scattering in PWZ (dash dotted line), and synchrotron radiation and IC emission from IBS (dotted line). As we can see, the steady component observed by \textit{Fermi}/LAT can be well fitted by the outer gap emission from the pulsar magnetosphere, while the boosted emission from the shock and IC scattering in the wind will also contribute to the $\gamma$-rays observed by \textit{Fermi}/LAT. We note that the predicted spectrum of the wind at SUPC is somehow higher than the observational data which may be due to our neglect of the Compton-drag effect. Finally, the orbital variations of $\gamma$-ray flux in 0.1-100GeV are presented in Fig. \ref{fig:GeV}.

\section{Conclusion and Discussion}\label{sect:summary}
Gamma-ray binaries are unique astrophysical laboratories for studying particle acceleration and physical properties of outflows from energetic pulsars and massive stars. The outer gap, the cold pulsar wind and the termination shock are the most likely regions that produce the observed $\gamma$-rays. We examined non-thermal emissions from the binary system with different shock structures and viewing angles, and applied the emission model to J1018. For IC emissions from the un-shocked wind, we demonstrated that the presence of the termination shock and the viewing angles determine the length of the un-shocked wind region towards the observer, and thus affect $\gamma$-ray flux.
Alternatively, the Doppler-boosting effect has a strong influence on the shock radiations. Depending on the relation between the shock structure and the viewing angle, two different patterns of X-ray/TeV light curves will be observed. In particular, when the orbital inclination angle $i$ is large enough, LOS will pass through the shock cone twice per orbit, and two sharp peaks will be observed around INFC, otherwise, only one or less spike can be observed in one orbit.
Under the pulsar scenario, we studied HE emissions from J1018. We show that the periodic sharp peaks around phase 0 in the keV/TeV light curves of J1018 are caused by a boosted emission from the shock around INFC, while other broad sinusoidal modulations likely originates from a deflected tail at a larger distance. The data analysis of \textit{Fermi}/LAT indicates that the 0.1-100 GeV flux contains a steady component that does not change with the motion of the pulsar, and a modulated component which manifests flux maximum around INFC. The steady component can be well fitted by the outer gap emissions from the pulsar magnetosphere, while the modulated component is caused by boosted emissions from the shock and IC emissions from the wind.
Our model basically agrees with a previous study by \cite{An2017}, which performed a more delicate analysis and modeling of J1018. Both our results suggest the compact object of J1018 is an energetic pulsar with $L_{\rm sd}\sim 10^{36}\ {\rm erg\cdot s^{-1}}$. Among the orbital parameters, the inclination angle of the orbit is the most uncertain one, which has a strong impact on the modeling. We adopt modest values of $i=45^{\circ}$ and $\eta=0.05$, which are close to those of \cite{An2017} with $i=50^{\circ}$ and $\eta=1/25$. Strictly speaking, the mass-loss rate and wind velocity of the companion star can be obtained via optical spectroscopy, therefore, the sharp flares of light curves at INFC can further constrain the properties of the compact object. These would be beneficial for the future search of pulsations from the putative pulsar.

Some other binaries display common features with those of J1018, such as LS 5039 and LMC P3. The most important characteristic of these systems is that all of them exhibit significant correlation between the keV and TeV flux, which indicates a common population of particles that emit these photons (\citealt{Zabalza2011}). Nevertheless, different from J1018, the GeV flux of LS 5039 manifests anti-correlation with keV/TeV light curves, which is unlikely to be produced by the shock radiations (\citealt{Takata2014,Chang2019}). Besides, the flux modulations of LS 5039 and LMC P3 are much smoother than that of J1018.
Since the orbits of LS 5039 and LMC P3 are more compact than J1018, the strong wind from the companion star will push the shock much closer to the pulsar with a comet-tail shock geometry rather than a hollow-cone, and LOS may be far away from the shock flows, which causes a more smooth variation of shock radiations.
Due to the limitations of the observational sensitivities, some system parameters of these binaries are not yet constrained well, such as the eccentricity $e$, the viewing angles $i,\omega$ and the properties of the compact objects. If the compact objects of these systems are rotation-powered pulsars, there will be significant emissions from the termination shock and the un-shocked wind region. Investigating the multi-wavelength emissions from these regions can provide further constraints on the system parameters and the related properties of massive companion stars and compact objects.

\begin{acknowledgements}
We are very grateful for the valuable suggestions and insightful comments from referees, which improved the manuscript significantly. We also thank Wang Hui-hui for the help on data analysis, and Prof. Cheng Kwong-sang for useful discussion on the manuscript.
J.T. is supported by the National Key R\&D Program of China under grant 2020YFC2201400 and the National Science Foundation of China under grant U1838102. Y.Y. is supported by the National Natural Science Foundation of China under grants 11822302 and 11833003. A.C. is funded by the China Postdoctoral Science Foundation under grant 2020M682392.

\end{acknowledgements}

\end{document}